\documentclass[prc,reprint,superscriptaddress,floatfix]{revtex4-1} %twocolumn reprint preprint

\usepackage{graphicx}
\usepackage{amsfonts}
\usepackage{amssymb}
\usepackage{amsmath}
\usepackage{epstopdf}
\usepackage{colortbl} %for the \color command - colored text
\usepackage{lineno} %for line numbers
\usepackage{setspace} %change line spacing
\newcommand{\un}[1]{\ensuremath{\ \mathrm{#1}}} %Use \un for units in math mode.
\usepackage{multirow}
\begin{document}

\title{Neutron lifetime measurements and effective spectral cleaning with an ultracold neutron trap using a vertical Halbach octupole permanent magnet array}
\author{K.~K.~H.~Leung}
\affiliation{Institut Laue-Langevin, BP 156, F-38042 Grenoble, France}
\affiliation{North Carolina State University, Raleigh, NC 27695, USA}
\affiliation{Triangle Universities Nuclear Laboratory, Durham, NC 27708, USA}
\author{P. Geltenbort} 
\affiliation{Institut Laue-Langevin, BP 156, F-38042 Grenoble, France}
\author{S.~Ivanov}
\affiliation{Institut Laue-Langevin, BP 156, F-38042 Grenoble, France}
\author{F.~Rosenau}
\affiliation{Institut Laue-Langevin, BP 156, F-38042 Grenoble, France}
\author{O.~Zimmer}
\affiliation{Institut Laue-Langevin, BP 156, F-38042 Grenoble, France}

\date{\today}

\begin{abstract}
Ultracold neutron (UCN) storage measurements were made in a trap constructed from a 1.3~T Halbach Octupole PErmanent (HOPE) magnet array aligned vertically, using the TES-port of the PF2 source at the Institut Laue-Langevin. A mechanical UCN valve at the bottom of the trap was used for filling and emptying. This valve was covered with Fomblin grease to induce non-specular reflections and was used in combination with a movable polyethylene UCN remover inserted from the top for cleaning of above-threshold UCNs. Loss due to UCN depolarization was suppressed with a minimum $2\un{mT}$ bias field. Without using the UCN remover, a total storage time constant of $(712 \pm 19)\un{s}$ was observed; with the remover inserted for 80~s and used at either 80~cm or 65~cm from the bottom of the trap, time constants of $(824 \pm 32)\un{s}$ and $(835 \pm 36)\un{s}$ were observed. Combining the latter two values, a neutron lifetime of $\tau_{\rm n} = (887 \pm 39)\un{s}$ is extracted after primarily correcting for losses at the UCN valve. The time constants of the UCN population during cleaning were observed and compared to calculations based on UCN kinetic theory as well as Monte-Carlo studies. These calculations are used to predict above-threshold populations of $\sim 5\%$, $\sim 0.5\%$ and $\sim 10^{-12}\%$ remaining after cleaning in the no remover, 80\un{cm} remover and 65\un{cm} remover measurements. Thus, by using a non-specular reflector covering the entire bottom of the trap and a remover at the top of the trap, we have established an effective cleaning procedure for removing a major systematic effect in high-precision $\tau_{\rm n}$ experiments with magnetically stored UCNs.
\end{abstract}

\maketitle

\section{Introduction \label{sec:introduction}}

The mean lifetime of a free neutron, $\tau_{\rm n}$, is a parameter of wide impact. It is used directly in calculations of the primordial helium abundance from big bang nucleosynthesis \cite{Coc2014, Iocco2009, Mathews2005}. In combination with measurements of neutron $\beta$-decay correlation coefficients, $\tau_{\rm n}$ can be used to determine the weak interaction vector and axial-vector coupling constants, $G_V$ and $G_A$. These constants are required for calculating solar and reactor neutrino fluxes and their detection efficiencies \cite{Mention2011, Zhang2013}, as well as in searches for beyond standard model scalar and tensor semi-leptonic charged currents, probing physics at energy scales beyond the TeV range \cite{Bhattacharya2012,Cirigliano2013,Gardner2013}. These motivations are described in the review papers on the field of neutron particle physics \cite{Abele2008a, Dubbers2011}, on neutron $\beta$-decay \cite{Nico2009}, and on $\tau_{\rm n}$ specifically \cite{Paul2009,Wietfeldt2011}, as well as in the conference proceedings \cite{neutronLifetime2014}.
  
Ultracold neutrons (UCNs) are defined as having kinetic energy, $E_{\rm k}$, below the neutron optical potential of well-chosen materials ($U_{\rm opt} \sim 100\un{neV}$) so that they can be stored for long periods of time in a ``bottle'', in principle limited only by $\tau_{\rm n}$. The idealized reflection probability $|R|^2$ off a material surface forming a step function of height $U_{\rm opt}$ is a function of $E_{\rm k \bot}$, the component of kinetic energy associated with the velocity component perpendicular to the surface. For $E_{ \rm k \bot} < U_{\rm opt}$, it is given by \cite{Golub1991}:
\begin{equation}
\label{eq:TheorylossprobperbounceElessV}
|R(E_{\rm k\bot})|^2 = 1 -2 f \sqrt{\frac{E_{\rm k\bot}}{U_{\rm opt}-E_{\rm k \bot}}}\;,
\end{equation}
where $f = W/U_{\rm opt}$, with $W$ being the imaginary part of the neutron optical potential. This expression is only valid for $f \ll 1$ and $U_{\rm opt} >0$.

UCNs stored in material bottles have been used in precise measurements of $\tau_{\rm n}$ \cite{Mampe1989a,Mampe1993a,Pichlmaier2010a, Arzumanov2000a, Serebrov2005, Serebrov2008}. However, full understanding of the interaction between a UCN and material surfaces has so far been elusive. For many decades, the observed $f$ were often 2--3 orders-of-magnitudes larger than predicted \cite{Golub1991}, which has been attributed the be caused by surface contamination. More recently, ``warming'' of UCNs ($\sim 10^{-5}\un{neV}$ average per reflection) has been observed as the source of the anomalous loss \cite{Geltenbort1999,Strelkov2000,Steyerl2002} with quasi-elastic scattering off capillary waves on liquid-walled surfaces or reflections from magnetic domains as the mechanism \cite{Lamoreaux2002,Serebrov2003}. However, to this day, measurements on solid surfaces with low contamination and at low temperatures still produce an-order-of-magnitude higher loss rates than expected \cite{Pokotilovski2005, Pokotilovski2006, Pokotilovski2008}.

To eliminate material losses when extracting $\tau_{\rm n}$ these experiments are usually performed with bottles with different volume-to-surface ratios, assuming the same coating properties, and then extrapolated to an infinite neutron mean free path. This method has been demonstrated to be difficult to do: an initial 5 standard deviation ($\sigma$) disagreement between the currently published most precise results, \cite{Arzumanov2000a} and \cite{Serebrov2005}, required a reanalysis of the former's systematic effects some 12 years later in order to reach agreement \cite{Arzumanov2012a}. Also, as pointed out recently, there is a 3.8 $\sigma$ disagreement between $\tau_{\rm n}$ measurements with material bottles and those done with cold neutron beams \cite{Byrne1996,Nico2005a, Yue2013}, suggesting there are unknown effects in either sets of experiments. This motivates the need for $\tau_{\rm n}$ measurements using magnetic traps.

Neutrons in a magnetic field $\vec{B}$ have a potential energy $U_{\rm mag} = -\vec{\mu}_{\rm n}\cdot\vec{B} = \pm\,(60.3\,\mathrm{neV\,T^{-1}}) |\vec{B}|$, where $\vec{\mu}_{\rm n}$ is the neutron's magnetic moment. Thus, it feels a Stern-Gerlach force, $\vec{F}_{\rm mag} = -\vec{\nabla }U_{\rm mag}$, allowing one spin-state, the ``low-field seekers'', to be reflected and confined by magnetic field gradients. The magnetic storage of UCNs was first proposed by \cite{Vladimirskii1961} some 50 years ago, and subsequently demonstrated in early experiments \cite{Abov1986,*Abov1986a,Abov1983,*Abov1983a, Paul1989}. The pioneering experiment of the current-era of UCN magnetic bottles is that of the NIST magnetic trap \cite{Huffman2000a,Brome2001}, which employed a superconducting quadrupole magnet combined with two end coils in an Ioffe trap configuration.

Due to the similar-sized effect of Earth's gravitational potential on the neutron, $U_{\rm grav} = (102\un{neV\,m^{-1}})\, h$, where $h$ is the height, a magneto-gravitational trap can be built for UCNs. This configuration has the advantage of being able to access UCNs from the top of the trap and has been demonstrated by \cite{Ezhov2009}, and more recently by the UCN$\tau$ experiment \cite{Walstrom2009, Salvat2014}, both employing permanent rare-earth magnets. There is also another proposed magneto-gravitational experiment using a large assembly of superconducting coils \cite{Materne2009}.

Above-threshold UCNs, neutrons with total energy $E_{\rm tot}$ above what is storable in a trap, $U_{\rm trap}$, can cause a decrease in the observed storage time away from $\tau_{\rm n}$. This is because such neutrons can exist in ``quasi-stable'' trajectories and survive for times comparable to $\tau_{\rm n}$. While this effect also exists in material traps, the typically more symmetrical design of magnetic traps and the lack of non-specular reflections from magnetic field gradients exasperates the problem. Therefore, it is critical that magnetic UCN bottle experiments employ non-specular reflections (to induce ``mode-mixing'') during the removal (or ``cleaning'') of above-threshold UCNs.

The effect of poor cleaning was studied in the NIST trap \cite{Coakley2005, OShaughnessy2009, Coakley2016}. The cleaning technique used was ramping down the radial magnetic field temporarily so that above-threshold UCNs collide with material on the trap sidewall. Detailed simulations show ramping to 30\% of the initial field is required to ensure above-threshold UCNs are sufficiently cleaned from the trap (to reduce the storage time shift to $< 1\un{s}$), but this reduces the initial number of well-trapped UCNs ($E_{\rm tot} < U_{\rm trap}$) to $30\%$ \cite{Coakley2016}.

In magneto-gravitational traps, a UCN remover can be inserted from the top of the trap, which reduces loss of well-trapped UCNs when cleaning. However, above-threshold UCNs should not exist in quasi-stable trajectories, otherwise they do not explore the trap volume efficiently and thus can take long times before colliding with the remover. UCN$\tau$ has an inherent asymmetry in its magnetic ``bowl'' design so that quasi-stable trajectories are reduced \cite{Berman2008,Walstrom2009}. Nevertheless, from UCN tracking Monte-Carlo (MC) studies with a remover spanning the entire top of the trap and inserted to a height of $\sim 42\un{cm}$ to 44\un{cm} from the bottom of their 50\un{cm} high bowl, it was found these above-threshold UCNs survived relatively long cleaning periods. Inclusion of field ripples, which come from the inherent discrete changes in the magnetization direction along a Halbach array, were required to greatly improve the cleaning time in the simulations \cite{Walstrom2009}.

It is the removal time of above-threshold UCNs with $E_{\rm tot}$ a few neV above $U_{\rm trap}$ that is the most critical for determining the effectiveness of a cleaning procedure. In general, there is a trade off between increasing the cleaning effectiveness with decreasing the number of well-trapped UCNs remaining. The latter depends on the initial UCN spectrum loaded and the effective volume of the trap (see Sec.~\ref{sec:experimentSetup}). Another important facet of a good cleaning procedure is to not warm well-trapped UCNs to above $U_{\rm trap}$ after cleaning. This can occur from magnetic field ramping or doppler reflections off moving surfaces.

Our present paper describes results from the 1st phase measurements performed between 2009--2011 using the Halbach Octupole array of PErmanent (``HOPE'') magnets \cite{Leung2013a,Leung2014}, which is a 1.2\un{m} long, has a inner bore radius $\rho_{\rm trap} = 46.8\un{mm}$, and a nominal magnetic flux density of $|\vec{B}(\rho_{\rm trap})| = 1.3\un{T}$ at the inner walls. In this 1st phase, the bottom superconducting coil required to remove all interactions with material during storage was not yet installed. Instead, a mechanical UCN piston valve is used to close off the trap. The primary measurements were performed on the TES-port of the PF2 UCN source at the Institut Laue-Langevin (ILL), Grenoble, France. The UCN flux from this port is $\sim 20$ times less than from the other ports of the PF2 source. 

The goal of these measurements was to demonstrate an effective technique for above-threshold UCN cleaning. The idea is to use the trap vertically and to employ a non-specular reflecting surface at the bottom of the trap combined with a remover at the top. Because of gravity all UCNs are guaranteed to make mode-mixing reflections from the bottom surface, thus eliminating quasi-stable trajectories.

The mechanical UCN valve in these 1st phase measurements offers an emptying time constant of $\sim 2\un{s}$, allowing the number of remaining UCNs in the trap to be studied during cleaning. In Sec.~\ref{sec:cleaning}, these observations are compared with kinetic theory calculations \footnote{Kinetic theory assumes UCNs are in ``mechanical equilibrium''; i.e., the ensemble of UCNs are described by kinetic gas theory (see \cite{Golub1991} for details). This property shall be simply referred to as ``kinetic theory'' in this paper.} and MC studies to model the UCN dynamics during cleaning. In Sec.~\ref{sec:lifetime}, we extract $\tau_{\rm n}$ from the different measurements to observe the effect of poor cleaning.

This cleaning scheme can be employed when the bottom superconducting coil is installed for the next phase, full 3D magnetic trap measurements (described in Sec.~\ref{sec:conclusion}). It is this work that led to the change from the previous proposed horizontal trap configuration \cite{Leung2009} to a vertical one \cite{Leung2014}. Besides for cleaning, a vertical magneto-gravitational configuration has other advantages, such as live monitoring of depolarized or quasi-elastically scattered (``warmed'') UCNs. It also removes unwanted trapping of neutron decay electrons \cite{Leung2013a}. 

Two other known systematic effects required to be overcome by UCN magnetic bottle $\tau_{\rm n}$ experiments are phase-space evolution and depolarization of stored UCNs. The former effect occurs if: (a) the ensemble of UCNs take long times to uniformly occupying the phase-space of the trap and (b) if the detection efficiency of the surviving UCNs or the decay products has a phase-space dependence. This effect will not be addressed further in this paper, besides by qualitatively stating that the cleaning procedure employed will reduce (a) by providing a high frequency of non-specular reflections during cleaning, and the vertical configuration where UCNs are emptied through a large hole relative to the trap volume will reduce (b). 

The depolarization of stored UCNs is caused by Majorana spin-flips (i.e. transitions of low-field seekers to high-field seeker) when the adiabatic condition, 
\begin{equation}
\frac{2|\vec{\mu}_{\rm n}\cdot \vec{B}|}{\hbar} \gg \frac{|{\rm d}\vec{B}/{\rm d}t|}{|\vec{B}|} = \vec{v}\cdot\frac{\vec{\nabla}|\vec{B}|}{|\vec{B}|} \;,
\end{equation}
is violated. This can occur at regions where the field is small or where the field changes rapidly. The calculated loss rate from this effect can vary by several orders-of-magnitude \cite{Walstrom2009, Steyerl2012a} and thus warrants experimental study. In our 1st phase experiments, a $\sim 2\un{mT}$ bias field is provided by a copper-wire solenoid to suppress depolarization. In Sec.~\ref{sec:depolarization}, experimental results studying depolarization by scanning the bias field are presented.

Because of these systematic effects, in order to understand the discrepancy amongst the bottled UCN $\tau_{\rm n}$ experiments and between these and beam $\tau_{\rm n}$ experiments, several robust $\sim$ 1-s-precision magnetic bottle UCN experiments will be required. The size of our trap (physical volume $\sim 2\un{L}$) is small, but it offers an excellent control of systematics and potentially a charged product detection scheme \cite{Leung2009, Rosenau2015}. The near-term goal of the HOPE project is to provide a ``1-s-precision'' measurement using the SUN-2 UCN source at the ILL \cite{Zimmer2011,Piegsa2014, Leung2016}. Such a precision is currently still sufficient for Big Bang Nucleosynthesis calculations. A ``sub-1-s-precision'' experiment might also be possible with HOPE but will require access to future UCN sources, such as the SuperSUN source \cite{Zimmer2015}. The work described here constitutes a significant milestone towards reaching these aims for the HOPE project. 

\section{Experimental setup \label{sec:experimentSetup}}

\begin{figure}[tb!]
\begin{center}
\includegraphics[width=2.7in]{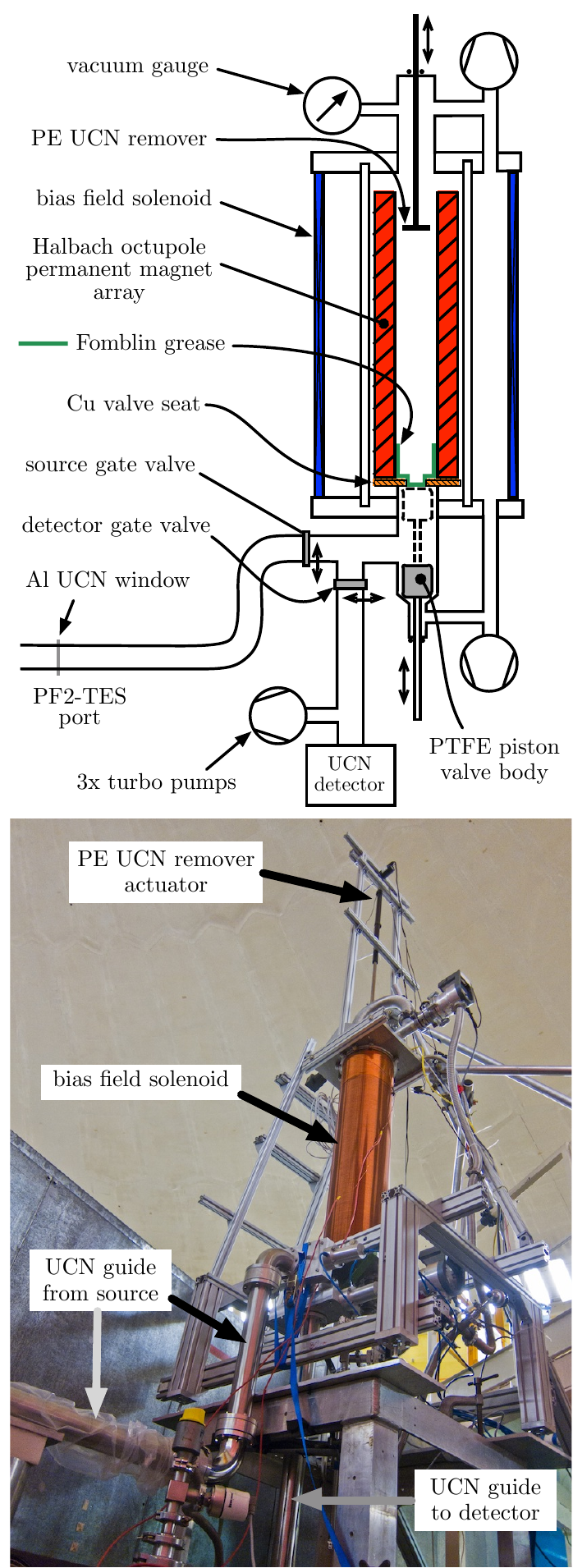}
\end{center}
\caption{(Color online) Schematic and photo of the 1st phase experiments with the HOPE magnetic trap performed on the PF2-TES port using a mechanical UCN piston valve for filling and emptying.}
\label{fig:setup}
\end{figure}

The 1st phase experiment setup used in 2011 is shown in Fig.~\ref{fig:setup}. The 1.2\un{m} long octupole array is orientated vertically. The mechanical UCN piston valve at the bottom of the trap is made of a copper valve seat with a 6\un{cm} opening and a movable PTFE piston valve body 6.2\un{cm} in diameter. The valve is opened until the piston is retracted pass the opening of a T-section of electropolished stainless steel ($U_{\rm SS} = 185\un{neV}$ and $f_{\rm SS} \approx 1\times10^{-4}$) UCN guide at the bottom of the trap.

The copper and PTFE surfaces exposed to UCNs are covered with Solvay Solexis RT-15 Fomblin grease ($U_{\rm Fomblin} \approx 108\un{neV}$ and $f_{\rm Fomblin} \approx 2\times10^{-5}$, and discussed in detail in Sec.~\ref{sec:lifetime}) applied by hand using powder-less nitrile gloves. This was done to reduce losses from surface impurities or leakages through small gaps at the seal of the valve. The grease does not appear flat visually and thus UCNs are expected to make non-specular reflections from this surface. Fomblin grease is also used to cover the bottom 8\un{cm} of the magnet inner bore. This was done because $|\vec{B}|$ drops at the ends of the magnets. From finite-element calculations and verified by flux probe measurements, at 8~cm from the ends $|\vec{B}| > 99.5\% $ of the field deep inside the array. The calculated neutron optical potential of the Nd$_2$Fe$_{14}$B magnet material is $U_{\rm NdFeB} = 170\un{neV}$ with $f_{\rm NdFeB}=1.7\times10^{-3}$. However, the inner bore is partly coated with epoxy from the manufacturer for mechanical reasons.

A polyethylene (PE) UCN remover plate ($U_{\rm PE} = -8.5\un{neV}$ and $W_{\rm PE} = 0.5\un{neV}$) with diameter 6.5\un{cm} and thickness 5\un{mm} is lowered into and raised out of the trap from the top. Its position is reproduced to within a few millimeter. The PE diameter is less than the magnet diameter (9.4\un{cm}) so that there's vacuum pumping of the trapping region even when the remover is lowered. The pressure is measured with a cold cathode gauge at the top of the trap. It is estimated to be $\sim 5 \times 10^{-5}\un{mbar}$ by temporarily closing the vacuum valves to the turbo pumps and observing the equilibrium value during the holding time. The pressure does not jump when the either the UCN valve or remover is actuated. The vacuum feedthroughs of these are made from double elastomer O-rings with active pumping to $\sim10^{-3}\un{mbar}$ in between.

The trap is installed approximately 5\un{m} from the aluminum front window ($U_{\rm Al} = 54\un{neV}$) of the PF2-TES port, connected using electropolished stainless steel UCN guides with internal diameter 66~mm. Using two 90$^\circ$ bends, the bottom of the trap is raised 60~cm above this window. The piston valve, mentioned earlier, as well as the source gate valve (GV) and the detector GV allow several procedures needed to perform storage experiments: filling of the trap, closing the trap, and emptying remaining UCNs to a detector. There is another GV upstream of the Al foil. These valves were driven pneumatically and have opening and closing times of 1--2\un{s}.

UCN detection is made with a $^3$He gaseous wire chamber positioned 1.5~m below the bottom of the trap so that UCNs gain sufficient kinetic energy to overcome the potential barrier of its front aluminium window.  A turbo-molecular pump is connected through a small side-port just above the detector so that pumping occurs even when the detector GV is closed. The ambient background rate after shielding the detector with polyethylene and boron rubber was $\approx 0.2\un{s^{-1}}$; without shielding it was $\approx 2\un{s^{-1}}$.

A bias field along the vertical axis, used to suppress UCN depolarization, is produced using a solenoid. This coil is made with 2-layers of insulated 1.4~mm diameter copper wire, and has a diameter of 25~cm and a length 5~cm longer than the 1.2~m magnet array. A bias field solenoid current of 4\un{A} was nominally used.

\begin{figure}[tb!!]
\begin{center}
\includegraphics[width=3.2in]{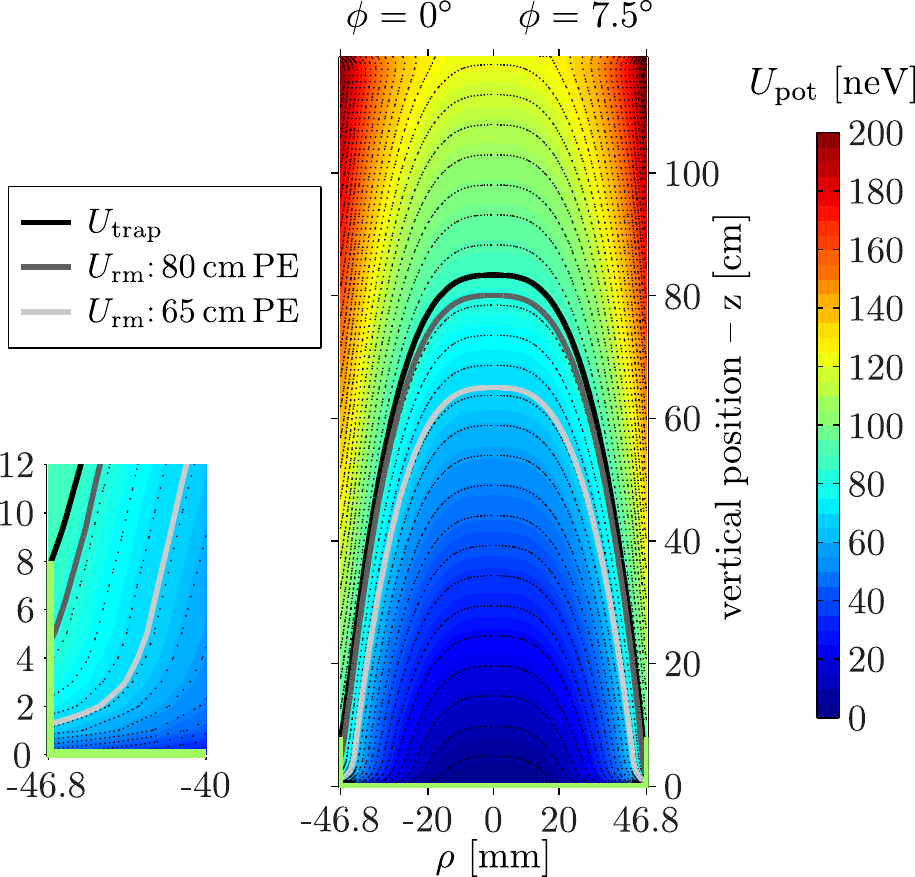}
\end{center}
\caption{(Color online) (Main plot) Contour plots of the UCN potential energy $U_{\rm pot}(\vec{r})$ in the trapping region on vertical slices passing through $\rho=0$. On the left for $\phi = 0$ and on the right for $\phi =7.5^{\circ}$ (see text). Zero potential is defined to be at $z = 0$ and $\rho=0$. The aperture for the UCN piston valve at the bottom of the trap and the epoxy on the magnet sidewall is not shown. The location of the Fomblin grease on the valve body, valve seat and magnet sidewall is shown with lines in a color to match that of $U_{\rm Fomblin}$. (Left) Zoomed into the bottom-left hand corner of the main plot (i.e. the $\phi = 0$ slice). The trapping potential $U_{\rm trap} = 84\un{neV}$, shown as the solid-black line, is defined by the height of the Fomblin grease on the sidewall. This contour is 5\un{cm} lower when $\phi=7.5^{\circ}$. The contours of the two remover potentials $U_{\rm rm}$ for $h_{\rm PE} = 80\un{cm}$ and 65\un{cm} are also shown. These contours are 2\un{cm} and 0.5\un{cm} lower when $\phi=7.5^{\circ}$, respectively.}
\label{fig:potentialMap}
\end{figure}

The potential energy $U_{\rm pot}(\vec{r})$ for a neutron at a position $\vec{r}$ is given by:
\begin{equation}
U_{\rm pot}(\vec{r}) = U_{\rm grav}(\vec{r}) + U_{\rm mag}(\vec{r}) + U_{\rm opt}(\vec{r})\;,
\end{equation}
where $U_{\rm pot}(\vec{r}) = 0$ is defined to be the minimum potential inside the trapping region. The total energy is thus given by $E_{\rm tot} = U_{\rm pot}(\vec{r}) + E_{\rm k}(\vec{r})$. A contour plot of $U_{\rm pot}(\vec{r})$ for our trapping configuration is shown in Fig.~\ref{fig:potentialMap}. For our trap, a cylindrical coordinate system with radial position $\rho = 0$ aligned with the central axis of the magnetic array is used. It should be noted that $|\vec{B}(\rho,\phi,z)|$, where $\phi$ is the azimuthal angle and $z$ the height, is not $\phi$-symmetric. It takes on its smallest value $|\vec{B}(\rho_{\rm trap},\phi=0)| = 1.3\un{T}$ and its largest value $|\vec{B}(\rho_{\rm trap},\phi=7.5^{\circ})| = 1.6\un{T}$ for $z$ more than a few centimeter inside the array. This pattern is repeated 16 times azimuthally. (A detailed description of the magnetic field is found in \cite{Leung2013a, Leung2014}.) The $\phi$-variation of $|\vec{B}|$ are included in the subsequent calculations.

\begin{figure}[tb!]
\begin{center}
\includegraphics[width=3in]{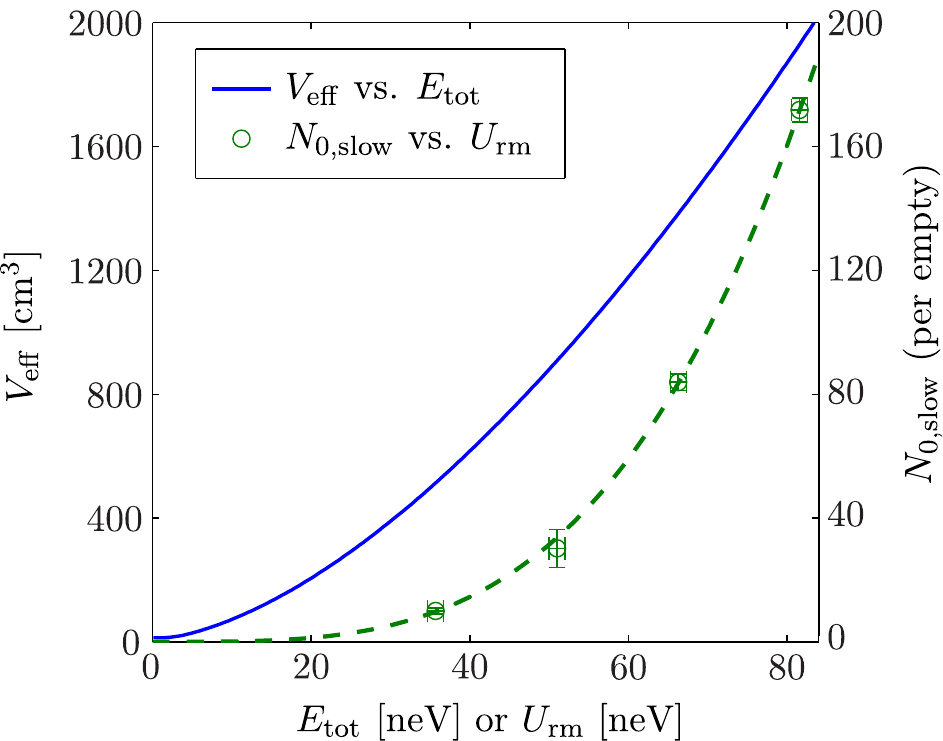}
\end{center}
\caption{(Color online) Blue-line: The effective volume $V_{\rm eff}$ of our trapping configuration versus the total energy $E_{\rm tot}$ for without the PE remover. When the remover is in-place, UCNs with $E_{\rm tot}>U_{\rm rm}$ experience a reduced $V_{\rm eff}$. Green circles with error bars: The measured $N_{\rm 0, slow}$, the number of UCNs initially loaded into the trap with $E_{\rm tot}<U_{\rm rm}$, for different PE remover heights. Green dashed line: the green points fitted with Eq.~\ref{eq:integralSpectrumFit} to determine the shape of the integral UCN spectrum $N_0(E_{\rm tot}< U_{\rm rm})$. The relationship between $V_{\rm eff}(E_{\rm tot})$ and $N_0(E_{\rm tot}< U_{\rm rm})$ is described in the text near Eq.~\ref{eq:integralSpectrumFit}.}
\label{fig:effectiveVolume}
\end{figure}

UCNs with increasing $E_{\rm tot}$ can occupy a greater volume of space. Because of this effect a quantity called the effective volume can be useful. Following the definition from \cite{Golub1991}, the effective volume is given by:
\begin{equation}
\label{eq:Veff}
V_{\rm eff}(E_{\rm tot}) =  \int_{V} \,\sqrt{\frac{ E_{\rm k}(\vec{r})}{E_{\rm tot}}} \, \:,
\end{equation}
which assumes kinetic theory. The integral is performed over only the volume accessible to the UCNs [i.e. $E_{\rm k}(\vec{r}) \geq 0$]. $V_{\rm eff}(E_{\rm tot})$ for our trapping configuration is shown in Fig.~\ref{fig:effectiveVolume}.

The effective volume naturally leads to the definition of a quantity that we shall call the effective loss area:
\begin{equation}
\label{eq:Aeffloss}
A_{\rm eff\,loss}(E_{\rm tot}) \equiv  {\rm Re}\left\{ \int_{S}\sqrt{\frac{E_{\rm k}(\vec{r})}{E_{\rm tot}}} \, \bar{\mu}[E_{\rm k}(\vec{r})] \, {\rm d}S \right\},
\end{equation}
where $E_{\rm k}(\vec{r})$ is taken to be at $\vec{r}$ just outside the surface element ${\rm d}S$ [i.e., $E_{\rm k}(\vec{r}) = E_{\rm tot} - U_{\rm grav}(\vec{r}) - U_{\rm mag}(\vec{r})$] and $\bar{\mu}$ is the average loss probability per reflection. This quantity, inspired by work in \cite{Pendlebury1994,Golub1991}, becomes useful in describing the loss of UCNs on a surface that is not at equal $U_{\rm grav}+U_{\rm mag}$, which is true for any surface that is not horizontal or existing in a magnetic field gradient. (E.g., the trap sidewall as well as the PE remover.) By taking the real component, only elements ${\rm d}S$ accessible by UCNs contribute to the integral.

\begin{figure}[tb!]
\begin{center}
\includegraphics[width=3in]{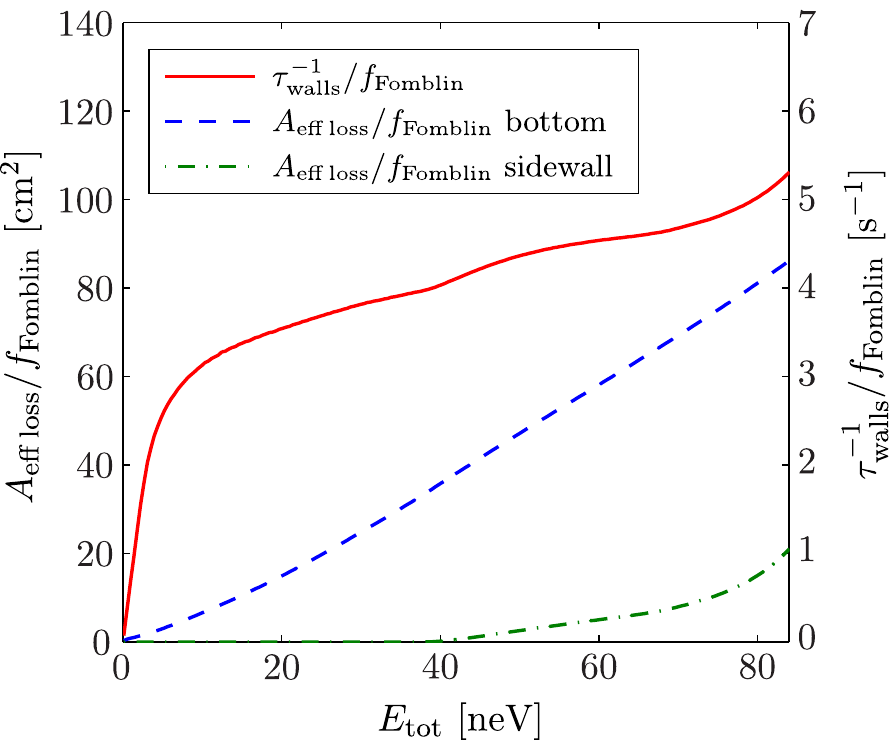}
\end{center}
\caption{(Color online) The effective loss area $A_{\rm eff\,loss}(E_{\rm tot})$ normalized by $f_{\rm Fomblin}$ for the Fomblin grease at the bottom and sidewall. The loss time constant due to both these surfaces, $\tau^{-1}_{\rm walls}$, also normalized by $f_{\rm Fomblin}$, is also shown. All these quantities assume kinetic theory.}
\label{fig:tauWalls}
\end{figure}

By assuming kinetic theory and integrating Eq.~(\ref{eq:TheorylossprobperbounceElessV}) over all incident angles, it can be derived to be \cite{Golub1991}:
\begin{eqnarray}
\bar{\mu}[E_{\rm k}(\vec{r})] = 2f\Biggl\{ \frac{U_{\rm opt}(\vec{r})}{E_{\rm k}(\vec{r})} \, \arcsin\!\left[ \frac{E_{\rm k}(\vec{r})}{U_{\rm opt}(\vec{r})}\right]^{\frac{1}{2}} \nonumber \\* - \left[ \frac{U_{\rm opt}(\vec{r})}{E_{\rm k}(\vec{r})} - 1\right]^{\frac{1}{2}}\Biggr\}\;,
\end{eqnarray}
where $U_{\rm opt}(\vec{r})$ is taken to be at the surface element ${\rm d}S$. This expression is valid for $f \ll 1$ and $U_{\rm opt} >0$. Plots of $A_{\rm eff\,loss}(E_{\rm tot})/f_{\rm Fomblin}$ for the Fomblin at the bottom and at the sidewall of the trap are shown in Fig.~\ref{fig:tauWalls}.

\begin{figure}[tb!]
\begin{center}
\includegraphics[width=3.0in]{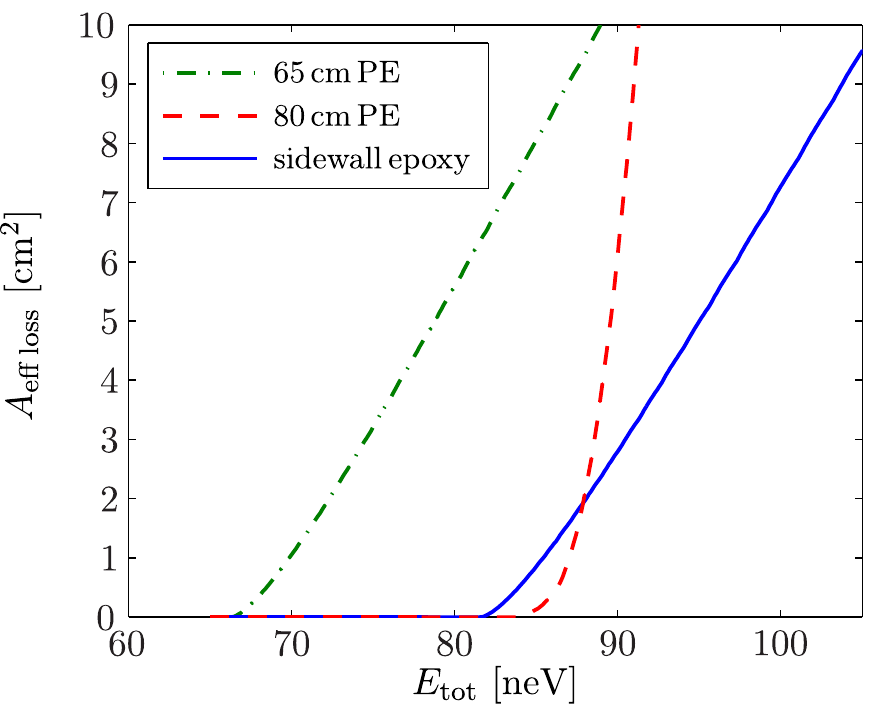}
\caption{(Color online) The effective loss area $A_{\rm eff\,loss}(E_{\rm tot})$ for the epoxy on the trap sidewall and for the PE UCN remover, placed at two different heights, $h_{\rm PE}$. The epoxy is assumed to have the same $U_{\rm opt}$ and $f$ as the PE.}
\label{fig:Aeffloss}
\end{center}
\end{figure}

The PE, as well as the hydrocarbon-based epoxy on the trap sidewall, which we shall treat as having the same $U_{\rm opt}$ and $f$, up-scatter UCNs to well above $U_{\rm trap}$ so they are lost. Since $U_{\rm opt} < 0$ and $W\approx 0$ for these materials, the previous expression is not valid. Instead, $\bar{\mu}$ for this class of materials, assuming kinetic theory, is \cite{Brose2012}:
\begin{equation}
\label{eq:PElossProbability}
\bar{\mu}(E_{\rm k}) = \frac{8E_{\rm k}^2 - 8\sqrt{E_{\rm k}}(E_{\rm k}-U_{\rm opt})^{\frac{3}{2}}-12E_{\rm k}U_{\rm opt} + 3U_{\rm opt}^2}{3U_{\rm opt}^2},
\end{equation}
where the $\vec{r}$ dependence of $E_{\rm k}(\vec{r})$ is not explicitly written. A plot of $A_{\rm eff\,loss}(E_{\rm tot})$ for the PE remover used at two different heights, $h_{\rm PE}$, and for the epoxy on the trap sidewall is shown in Fig.~\ref{fig:Aeffloss}.

The trapping potential $U_{\rm trap}$ for this configuration is defined by when UCNs start interacting with epoxy on the sidewall (see Fig.~\ref{fig:potentialMap}), since this is when UCNs get lost quickly (compare Figs.~\ref{fig:tauWalls} and \ref{fig:Aeffloss}). Its value is $U_{\rm trap} = 84\un{neV}$ and is determined by the height of the Fomblin on the sidewall. The cleaning cut-off potential is denoted by $U_{\rm rm}$. When no remover is used $U_{\rm rm} = U_{\rm trap}$. When the PE remover is at a height $h_{\rm PE}$ measured from the bottom of the trap, $U_{\rm rm}$ is the smallest $E_{\rm tot}$ when $A_{\rm eff\,loss}> 0$. These values for when $h_{\rm PE}  = 65\un{cm} \text{ and } 80\un{cm}$ for the storage measurements are $U_{\rm rm} = 81.6\un{neV}$ and $66.3\un{neV}$, respectively.

\section{Measurements \label{sec:results}}
The sequence of steps for each UCN storage measurement is as follows: \\
(1) open piston valve, open source GV, and close detector GV -- the start of the 100\un{s} long period of filling the trap with UCNs from source; \\
(2) start lowering PE remover -- at $1\un{cm\,s^{-1}}$ with remover initially located 10\un{cm} above its lowered position $h_{\rm PE}$;\\
(3) close piston valve -- define as $t=0$;\\
(4) wait 5\un{s}, close source GV and open detector GV; \\
(5) at $t = 80\un{s} \equiv t_{\rm clean}$ begin raising the remover -- at $1\un{cm\,s^{-1}}$ for 10\un{cm};\\
(6) at $t=t_{\rm hold}$ open piston valve -- empties UCNs remaining in trap to UCN detector.\\
To observe the number of UCNs remaining in the trap during the cleaning, (5) can occur before (4) (i.e. $t_{\rm hold} < t_{\rm clean}$). The shortest $t_{\rm hold}$ used was 20\un{s}.

\begin{figure}[tb!]
\begin{center}
\includegraphics[width=3.2in]{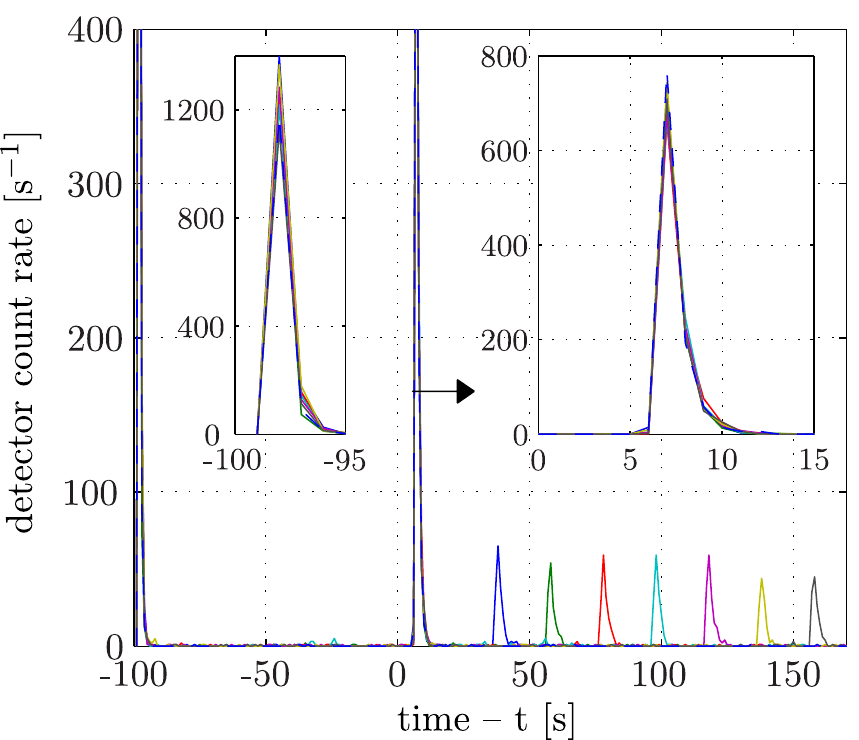}
\end{center}
\caption{(Color online) The count rate at the UCN detector during several measurement procedures for $t_{\rm hold} \geq 35\un{s}$. The details of a procedure and the cause of the peaks are described in the text.}
\label{fig:exampleData}
\end{figure}

A plot of the detected count rate in 1\un{s} bins during the procedure repeated several times for varying $t_{\rm hold}$ is shown in Fig.~\ref{fig:exampleData}. The peak in the count rate at $-100\un{s}$ comes from the valve switching in step (1), which temporarily has the source GV and detector GV opened simultaneously. The peak beginning at 5\un{s} comes from emptying the UCNs trapped between the source GV and piston valve into the UCN detector when the detector GV is opened in step (4).

The peaks seen from 35\un{s} onwards are those of surviving UCNs in the trap emptied by step (6) after various $t_{\rm hold}$. The decay time constant of these emptying peaks are $1\text{--}2\un{s}$. This compares well with kinetic theory: a $ 2\un{s}$ time constant is expected for $E_{\rm tot} = 10\un{neV}$ emptying through the 6\un{cm} diameter piston valve opening from a 2\un{L} volume. The total number of surviving UCNs is calculated by summing counts from $t_{\rm hold}$ to $t_{\rm hold} + 60\un{s}$.

The background rate used to correct each measurement of the total number of surviving UCNs is calculated by using the last 40\un{s} of each emptying period (i.e. $> t_{\rm hold}+60\un{s}$) and the 30\un{s} before the emptying begins (when $t_{\rm hold}$ is sufficiently long). When combining data from different fill-and-empty procedures to form a storage curve, the background corrected counts are normalized to the ILL reactor power ($< 3\%$ change for all the data). The stability of the UCN flux from the PF2 reactor-based turbine source is sufficient for the level of precision required in these measurements. Furthermore, $t_{\rm hold}$ of storage procedures were varied in a pseudo-random way to mitigate the effects of filling variations.

UCN storage curves were measured for $h_{\rm PE}$ = 35\un{cm}, 50\un{cm}, 65\un{cm} and 80\un{cm}, as well as with no remover in place at all. The 35\un{cm} and 50\un{cm} measurements have too low statistics for detailed analysis of the cleaning and long storage times, however, they are useful for determining the UCN spectrum initially loaded into the trap.

\begin{figure}[tb!]
\begin{center}
\includegraphics[width=3in]{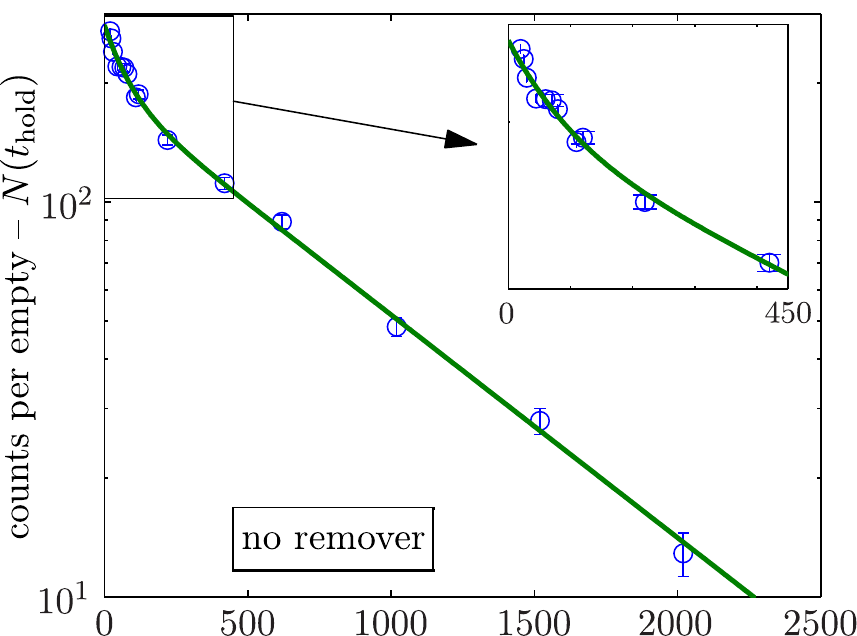}
\includegraphics[width=3in]{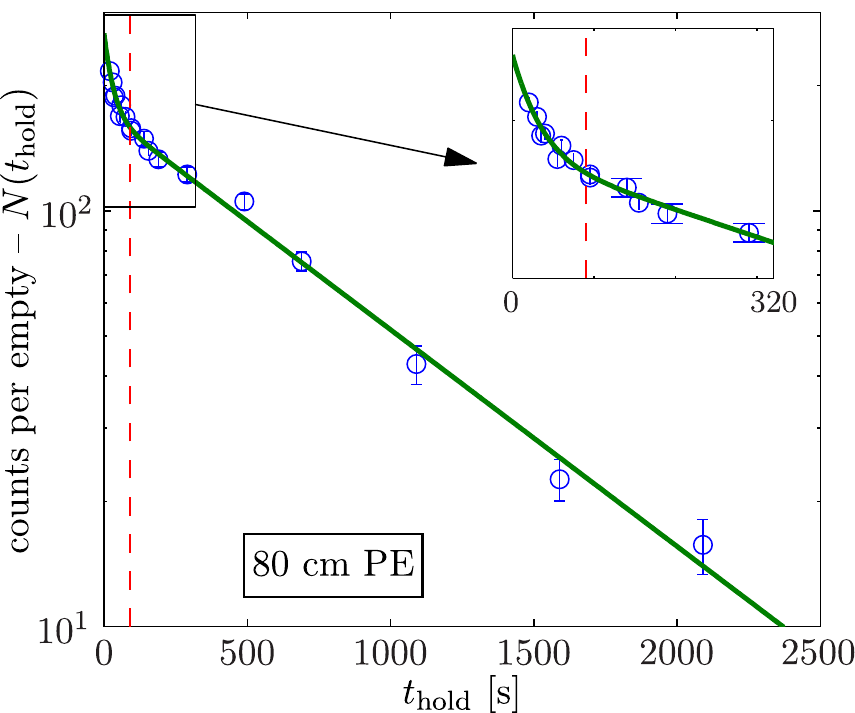}
\end{center}
\caption{(Color online) Storage curves for the no remover and $h_{\rm PE} = 80 \un{cm}$ measurements. The vertical dashed line is shown to indicate $t_{\rm clean} = 80\un{s}$, the time when the PE remover was raised. The zoomed-in regions are to show details for short $t_{\rm hold}$. The solid lines are from fits with a sum of two exponential decays (see text).}
\label{fig:storageCurve80cm}
\end{figure}

The storage curve for 80~cm PE and no remover measurements are shown in Fig.~\ref{fig:storageCurve80cm}. There is a quickly decaying component when $t_{\rm hold} < t_{\rm clean}$, indicating UCNs with $E_{\rm tot}>U_{\rm rm}$ are being cleaned out at this stage. At $t_{\rm hold} > t_{\rm clean}$, with the absorber being raised or starting to be raised, the decaying slows down and takes on a single exponential behavior. This same behavior is seen in the other storage curves.

The fitted curve to the data (as shown on Fig.~\ref{fig:storageCurve80cm}) is a sum of two exponential decays, i.e.,
\begin{equation}
N(t_{\rm hold}) = N_{\rm 0,\, fast}\,{\rm e}^{-t_{\rm hold}/\tau_{\rm fast}} + N_{\rm 0,\,slow}\,{\rm e}^{-t_{\rm hold}/\tau_{\rm slow}}\;.
\label{eq:doubleExpDecay}
\end{equation}
The best-fit parameters extracted from the no PE remover, 80~cm PE remover and 65~cm PE remover storage curves are shown in Table~\ref{tab:twoExpDecays} \footnote{The values for 80~cm PE are different to those in \cite{Leung2013a} because a point at $t_{\rm hold} = 15\un{s}$ has been discarded in the analysis of this paper so that for all three curves the shortest $t_{\rm hold} = 20\un{s}$ is used.}. $N_{\rm 0,\, fast}$ and $\tau_{\rm fast}$ are used to model the initial number and time behavior of UCNs with $E_{\rm tot} > U_{\rm rm}$, and $N_{\rm 0,\, long}$ and $\tau_{\rm long}$ for UCNs with $E_{\rm tot} < U_{\rm rm}$. In reality, the time-dependence contains a continuum of time constants that vary with $E_{\rm tot}$. This is discussed in Sections \ref{sec:cleaning} and \ref{sec:lifetime}. 

%These fits are just approximations, however. In reality, UCNs with $E_{\rm tot} < U_{\rm PE}$ have a continuum of $E_{\rm tot}$ dependent storage times given by $\tau^{-1}_{\rm n} + \tau^{-1}_{\rm walls}(E)$, and when the absorber is down, UCNs with $E > U_{\rm abs}$ have an storage time $\tau^{-1}_{\rm n} + \tau^{-1}_{\rm walls}(E_{\rm tot}) + \tau^{-1}_{\rm clean}(E_{\rm tot}) \approx \tau^{-1}_{\rm clean}(E_{\rm tot})$. As we shall see in Sec.~\ref{sec:cleaning}, $\tau^{-1}_{\rm clean}(E_{\rm tot})$ has a strong $E_{\rm tot}$-dependence. And in Sec.~\ref{sec:lifetime} it will be shown that $\tau^{-1}_{\rm walls}(E)$ is small and has a weak $E_{\rm tot}$-dependence so that a single exponential decay is sufficient for fitting the decay of UCNs with $E_{\rm tot} < U_{\rm abs}$. Because of this $\tau_{\rm n}$ can be extracted from the data.

\begin{table}

\begin{center}
\caption{The fitted-parameters of the double exponential decay function (Eq.~\ref{eq:doubleExpDecay}) used to model the decay curves for the no remover, 80~cm PE remover and 65~cm PE remover measurements.}
\label{tab:twoExpDecays}
\begin{tabular}{c c c c c c c}
\hline\hline
	          remover & $N_{\rm 0,\,fast}$ & $\tau_{\rm fast}$ & $N_{\rm 0,\, slow}$ & $\tau_{\rm slow}$ & $\chi^2_{\nu}$ \\ \hline
no       & $90 \pm 8$    &  $(80 \pm 16)\un{s}$ &$190 \pm 8$ & $(769 \pm 32)\un{s}$ & 2.1    \\
80~cm PE  & $95 \pm 17$&  $(30 \pm 7)\un{s}$ &  $172 \pm 4$ & $(833 \pm 33)\un{s}$ & 1.1  \\
65~cm PE & $72 \pm 10$&  $(37 \pm 7)\un{s}$ &  $84 \pm 3$ & $(855 \pm 38)\un{s}$ & 2.4  \\\hline \hline
\end{tabular}
\end{center}
\end{table}

The $N_{\rm 0,slow}$ values from these fit are a good estimate of the number of UCNs with $E < U_{\rm rm}$ loaded into the trap at $t_{\rm hold} = 0$, which will be denoted by $N_0(E_{\rm tot}<U_{\rm rm})$. These values from the 80\un{cm} and 65\un{cm} remover measurements, along with the values from 35\un{cm} and 50\un{cm} remover measurements, are shown in Fig.~\ref{fig:effectiveVolume}. In order to determine the form of the integral UCN spectrum, these points are fitted with the ansatz function:
\begin{equation}
N_0(E_{\rm tot}<U_{\rm rm}) = a\,U_{\rm rm}^x \;,
\label{eq:integralSpectrumFit}
\end{equation}
where $x$ and $a$ are the varied parameters. Including a conservatively estimated uncertainty due to the remover height of $\pm 1\un{neV}$, the extracted values are: $x = 3.50 \pm 0.16$ and $a = (4.0 \pm 2.8) \times10^{-5}\;{\rm neV}^{-x}$, with a $\chi^2_{\nu} = 0.18$.

The differential UCN spectrum $n_0(E_{\rm tot})$, the number of UCNs with energies in the interval $(E_{\rm tot},E_{\rm tot}+{\rm d}E)$, can be deduced to be $n_0(E_{\rm tot}) \propto E_{\rm tot}^{2.50\pm0.16}$ from the fit. From \cite{Golub1991}, we expect $n_0(E_{\rm tot}) \propto d_{\rm source}(E_{\rm tot})\,\epsilon(E_{\rm tot})\, V_{\rm eff}(E_{\rm tot})$, where $d_{\rm source}(E_{\rm tot})$ is the differential density of UCNs provided from the source and $\epsilon(E_{\rm tot})$ is an energy-dependent factor taking into account the transport efficiency from the source to the trap. The UCN PF2 turbine is expected to produce a constant phase-space density (i.e. Maxwellian) UCN spectrum, i.e., $d_{\rm source} \propto E_{\rm tot}^{0.5}$. The lowest energy UCNs that make it to the trap (i.e. $E_{\rm tot} \approx 0$) have a kinetic energy of $\approx 61\un{neV}$ just after the Al foil. Therefore, the transmission loss through the foil is expected to be $\propto 1/v$; i.e., $\epsilon(E_{\rm tot}) \propto E_{\rm tot}^{0.5}$. Finally, a fit of the calculated $V_{\rm eff}(E_{\rm tot})$ with the function $bE_{\rm tot}^y$, with $b$ and $y$ as the free-parameters, approximates $V_{\rm eff}$ well and yields $y = 1.6$, i.e., $V_{\rm eff}(E_{\rm tot}) \propto E_{\rm tot}^{1.6}$. Combining these, we see that $n_0(E_{\rm tot})$ agrees with what is expected.

\begin{figure}[tb!]
\begin{center}
\includegraphics[width=3in]{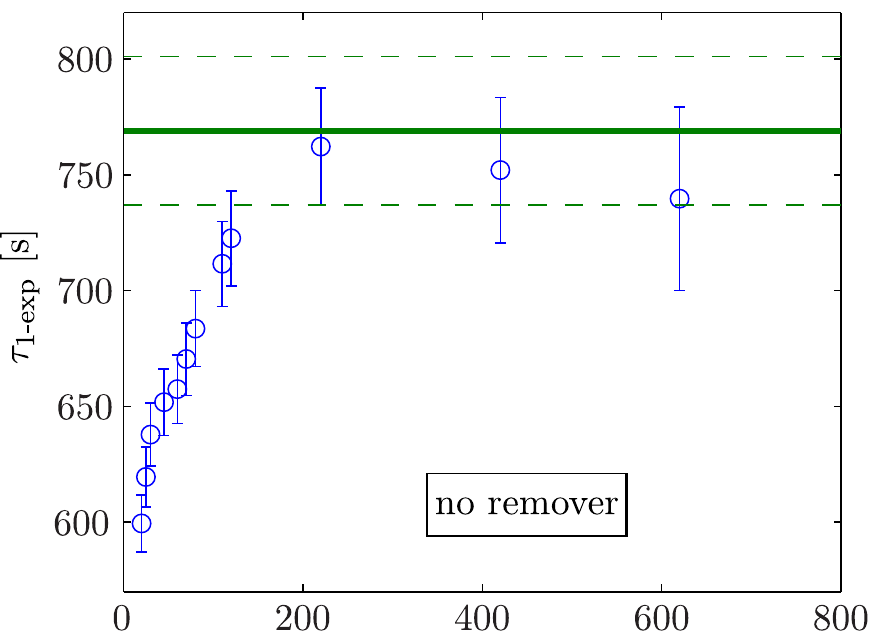}
\includegraphics[width=3in]{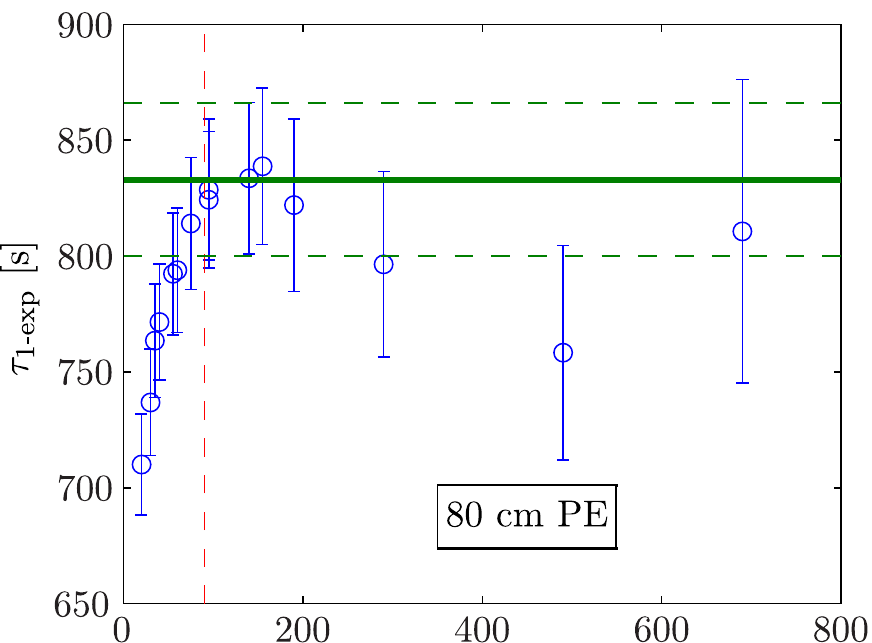}
\includegraphics[width=3in]{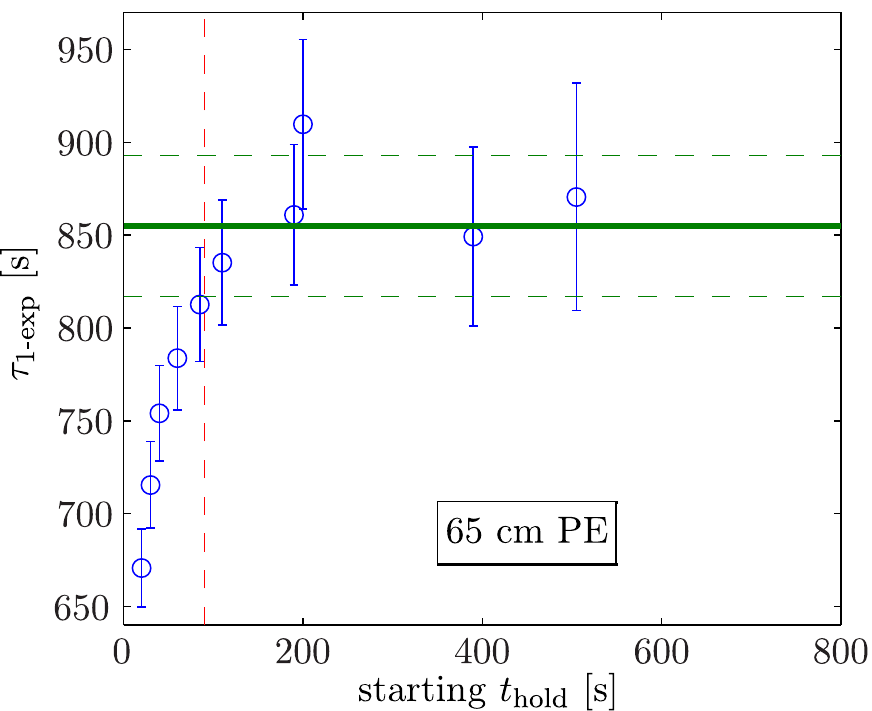}
\end{center}
\caption{(Color online) Analyses of the storage curves by fitting with a single exponential decay with time constant $\tau_{\rm 1\text{-}exp}$ and scanning the starting $t_{\rm hold}$ of the fit. The vertical dashed line is used to indicate $t_{\rm clean} = 80\un{s}$, the time when the remover was raised. For comparison, the $\tau_{\rm slow}$ values extracted from the sum of two exponential decay fits are also shown in these plots as horizontal green lines, with the dashed lines being the $\pm 1\sigma$ range.}
\label{fig:singleExpFits}
\end{figure}

An alternative way of analyzing the decay curves is to fit with a single decaying exponential with the starting $t_{\rm hold}$ of the fit delayed. If the initial fast decaying component is affecting the single exponential fits of the slow decay component, then the extracted time constant, $\tau_{\rm 1\text{-}exp}$, will be systematically shifted to shorter values. Plots using this analysis for the no remover, 80\un{cm} and 65\un{cm} remover storage curves are shown in Fig.~\ref{fig:singleExpFits}.

It can be seen that when no remover is used, one has to wait $\sim 200\un{s}$ before the quickly decaying component no longer affects the extracted $\tau_{\rm 1\text{-}exp}$ within the $\sim \pm 25\un{s}$ statistical error bars. If the data had better precision, then this could be even longer. For the storage curves where the remover is used, it can be seen that immediately after the absorber is raised at 90\un{s}, the $\tau_{\rm 1\text{-}exp}$ values are stable and are thus statistically unaffected by cleaning of $E_{\rm tot}>U_{\rm rm}$ UCNs. This is a good way of demonstrating the effectiveness of the cleaning procedure. In the next section, we try to understanding the cleaning process more, and in Sec.~\ref{sec:lifetime} we extract $\tau_{\rm n}$ from the storage time after $t_{\rm clean}$.

\section{Interpretation of trap cleaning \label{sec:cleaning}}

\begin{figure}[tb!]
\begin{center}
\includegraphics[width=3.2in]{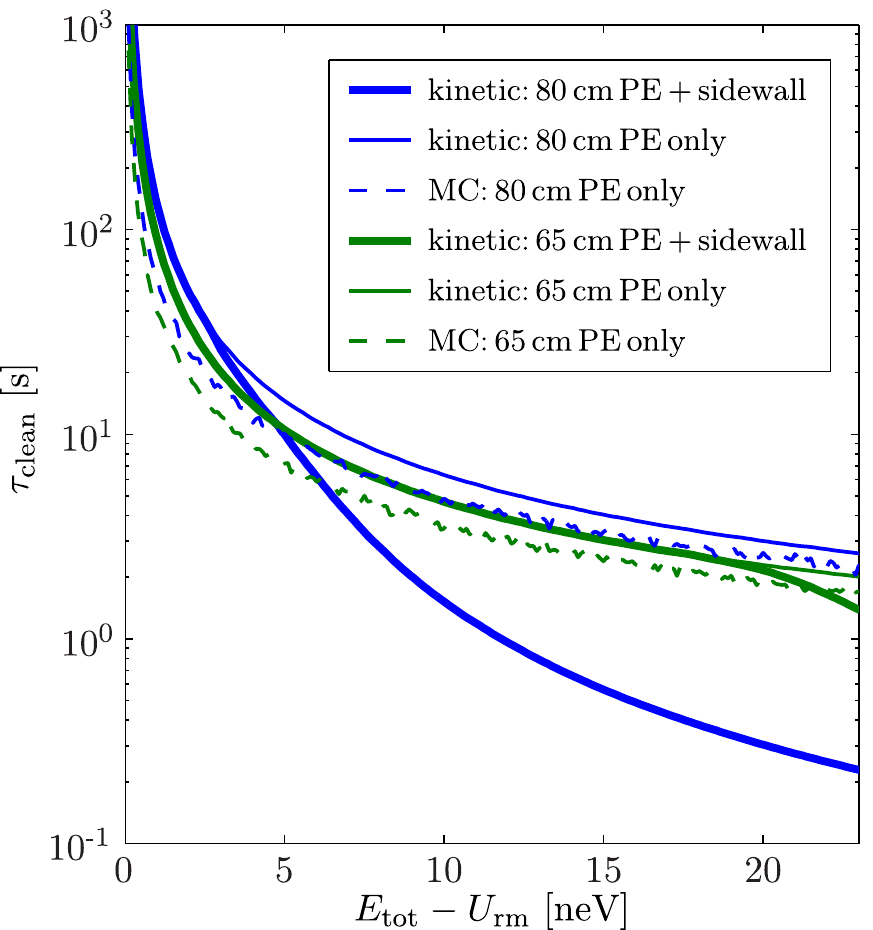}
\caption{(Color online) The calculated cleaning time constants $\tau_{\rm clean}(E_{\rm tot})$ using kinetic theory (Eq.~\ref{eq:tauCleanUCNgas}) and Monte-Carlo (MC) simulations. There are no losses at the sidewall in the MC simulations hence the kinetic theory calculations without sidewall loss are also plotted for comparison.}
\label{fig:cleaningAreasTimes}
\end{center}
\end{figure}

If kinetic theory is assumed, then the time constant of a UCN being lost on a surface is given by:
\begin{equation}
%\tau_{\rm clean}(E) = \frac{ 4 \sqrt{m}\,V_{\rm eff}(E_{\rm tot})}{ \sqrt{2E_{\rm tot}}\,A_{\rm eff\,loss}(E_{\rm tot}) } \; ,
\tau_{\rm clean}(E_{\rm tot}) = \sqrt{\frac{8m}{E_{\rm tot}}} \frac{V_{\rm eff}(E_{\rm tot})}{A_{\rm eff\,loss}(E_{\rm tot}) } \; ,
\label{eq:tauCleanUCNgas}
\end{equation}
where $m$ is the neutron mass and $A_{\rm eff\,loss}(E_{\rm tot})$ is the effective loss area of that surface (see Figs.~\ref{fig:tauWalls} and \ref{fig:Aeffloss}). When no absorber is used, above-threshold UCNs with $E_{\rm tot}>U_{\rm trap}$ experience loss caused by the sidewall only. The effective volume $V_{\rm eff}(E_{\rm tot})$ when the remover is in place is different to that shown in Fig.~\ref{fig:effectiveVolume}. UCNs with $E_{\rm tot}>U_{\rm rm}$ cannot access the volume behind the absorber, therefore the vertical upper integration limit of Eq.~\ref{eq:Veff} only goes up to $h_{\rm PE}$. The calculated $\tau_{\rm clean}(E_{\rm tot})$ for $E_{\rm tot} > U_{\rm rm}$ is shown in Fig.~\ref{fig:cleaningAreasTimes}. 

\begin{figure}[tb!]
\begin{center}
\includegraphics[width=3.2in]{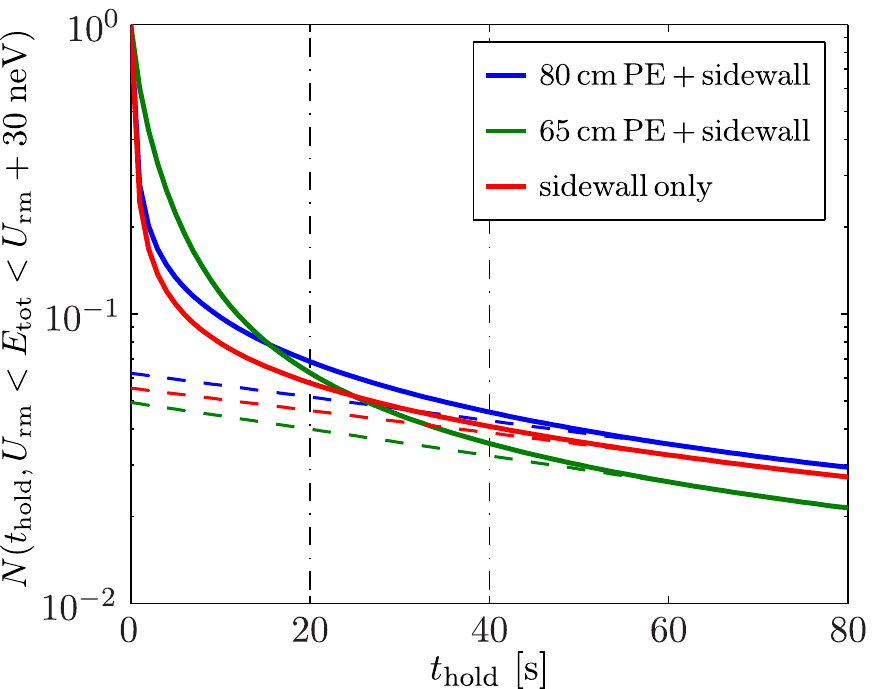}
\caption{(Color online) Calculated relative number of UCNs with energy $U_{\rm rm}<E_{\rm tot}<U_{\rm rm}+30\un{neV}$, denoted by $N(t_{\rm hold},U_{\rm rm}<E_{\rm tot}<U_{\rm rm}+30\un{neV})$, remaining in the trap during cleaning (i.e. when $t_{\rm hold} < t_{\rm clean} = 80\un{s}$) using $\tau_{\rm clean}(E_{\rm tot})$ from kinetic theory for the PE remover at 80\un{cm} and 65\un{cm} including the sidewall loss, and for no remover (i.e. sidewall loss only). The measured $E_{\rm tot}^{2.5}$ UCN spectrum is used in these calculations. The numbers are normalized to unity at $t_{\rm hold} = 0$. The dashed lines are single exponential fits between $20\un{s} < t_{\rm hold} < 40\un{s}$ (fit region indicated by the vertical dash-dotted lines) used to estimate $\tau_{\rm fast}$.}
\label{fig:simulatedCleaning}
\end{center}
\end{figure}

In order to link the calculated $\tau_{\rm clean}(E_{\rm tot})$ with the experimentally observed $\tau_{\rm fast}$, we calculate the number of UCNs with $E_1<E_{\rm tot}<E_1+\Delta E$ remaining in the trap: 
\begin{eqnarray}
N(t_{\rm hold},E_1<E_{\rm tot}< E_1+\Delta E) \nonumber \\*= \int^{E_1+\Delta E}_{E_1} n_0(E_{\rm tot}) \exp\left[\frac{-t_{\rm hold}}{\tau_{\rm clean}(E_{\rm tot})}\right]   {\rm d}E_{\rm tot} \;.
\end{eqnarray}
We do this for $t_{\rm hold}<t_{\rm clean}$ and for the energy range $U_{\rm rm}<E_{\rm tot}<U_{\rm rm}+30\un{neV}$ (see Fig.~\ref{fig:simulatedCleaning}). The measured UCN spectrum from Eq.~\ref{eq:integralSpectrumFit} is used for these calculations, assuming that the expression is still valid up to $E_{\rm tot} = 114\un{neV}$. UCNs with $E_{\rm tot}> U_{\rm rm}+30\un{neV}$ are lost very quickly and do not contribute to the UCNs surviving after $t_{\rm hold} = 20\un{s}$, the shortest $t_{\rm hold}$ used. Therefore, increasing $\Delta E$ for the plot further does not have an effect on the experimentally observable UCNs in these measurements.

The extracted value of $\tau_{\rm fast}$ is dominated by the $t_{\rm hold}$ points between 20\un{s} and 40\un{s} (Fig.~\ref{fig:storageCurve80cm}), where the 20\un{s} value is due to the shortest $t_{\rm hold}$ we used for our measurements. Therefore, to estimate the $\tau_{\rm fast}$ that we would observe from the calculated $N(t_{\rm hold})$, we fit with a single exponential between $20\un{s} < t_{\rm hold} < 40\un{s}$. The $\tau_{\rm fast}$ based on kinetic theory calculated this way are denoted by $\tau_{\rm fast,kinetic}$ and shown in Table~\ref{tab:tauFast}.

The PF2 UCN source does not provide pre-polarized UCNs, therefore the cleaning times of high-field seekers loaded into the trap are also studied. For high-field seekers the sign of $U_{\rm mag}$ is opposite to that of low-field seekers. This causes the region where $U_{\rm pot} = 0$ to be a ring with diameter of the magnet bore located at $z = 4\un{cm}$ (due to the drop in magnetic field at the end of the array). The threshold energy before high-field seekers start interacting with the epoxy on the magnet sidewall is $4\un{neV}$. UCNs can only enter the trap at $\rho < 30\un{mm}$ because of the aperture diameter of the Cu valve seat. Thus, only high-field seekers with $E_{\rm tot} > 62\un{neV}$ can enter the trap, and they are accelerated by the magnetic field gradient before colliding with the sidewall. Using Eq.~\ref{eq:tauCleanUCNgas}, all high-field seekers are found to have $\tau_{\rm clean} \approx 0.01\un{s}$. Therefore, they are expected to be cleaned from the trap so quickly that they do not affect the storage curves.

Thus far in the analysis of the cleaning we have assumed kinetic theory. This is only valid after UCNs make several non-specular reflections. Despite the Fombin grease at the bottom of the trap, at short $t_{\rm hold}$ kinetic theory will not be valid. To study this effect, a Monte-Carlo simulation of the cleaning time of $E_{\rm tot}>U_{\rm rm}$ UCNs in a simplified trap configuration was performed.

In this study, the trap sidewall is treated to be lossless and to provide specular reflections only. A lossless and 100\% diffuse reflecting surface is placed at $z = 0$, the position where UCNs start in the simulation. A vertical velocity component $v_z$ value is calculated from a specified $E_{\rm tot}$ and a randomly generated $\theta$ obeying Lambert's cosine law, $2 \pi \sin\theta \,\cos\theta\, {\rm d}\theta$. If $v_z^2/2 < gh_{\rm PE}$, where $g = 9.8\un{m\,s^{-1}}$, the UCN does not reach the remover at $h_{\rm PE}$. The time taken for the UCN to return to the bottom reflector, $2 v_z/g$, is added to the survival time of the UCN. If $v_z^2/2 > g h_{\rm PE}$, then the UCN reaches the remover and $(v_z-\sqrt{v_z^2-2gh_{\rm PE}})/g$ is added to the survival time. The loss probability is calculated using the equations for PE derived in \cite{Brose2012} (i.e., the $E_{\rm k \bot}$ version of Eq.~\ref{eq:PElossProbability}). If the UCN is lost then the simulation is terminated and its survival time recorded. If it is reflected, then a specular reflection is assumed to have occurred and the time taken for the UCN to reach $z=0$ is added to the survival time, and the next reflection from the bottom repeated.

This was done for $h_{\rm PE} = 65\un{cm}$ and 80\un{cm}. For every $E_{\rm tot}>U_{\rm rm}$, which are in 0.1\un{neV} bins, the survival times of 1000 UCNs were calculated. A histogram of these times was found to be well approximated by a single exponential decay. The time constant of this fitted decay is used to give $\tau_{\rm clean}(E_{\rm tot})$. The results of these MC calculations are also shown in Fig.~\ref{fig:cleaningAreasTimes}. Since there are no losses on the side-walls in the MC calculations, for a better comparison with the kinetic theory results, $\tau_{\rm clean}(E_{\rm tot})$ calculated with only the remover are also shown on the plot. The calculated $\tau_{\rm fast,\,MC}$ values, done in the same way as for $\tau_{\rm fast,\,kinetic}$, are shown in Table~\ref{tab:tauFast}.

Several things to note about the MC calculations: (1) Lambertian diffuse reflection is just a model for non-specularly reflection. If the diffuse probability is below 100\% then $\tau_{\rm clean}$ will scale linearly. (2) The MC calculations do not include the effects of the magnetic field gradient. The effect of the magnetic field in kinetic theory is to reduce the $A_{\rm eff\,loss}(E_{\rm tot})$ of the remover at large radii, which therefore increases $\tau_{\rm clean}(E_{\rm tot})$. When comparing the MC results with kinetic theory without the magnetic field, $\tau_{\rm clean}(E_{\rm tot})$ is $\sim 6$ times shorter when $E-U_{\rm rm} \rightarrow 0$, and they are approximately the same when $E-U_{\rm rm} \rightarrow 35\un{neV}$ \cite{Leung2013a}.

\begin{table}
\begin{center}
\caption{The experimentally observed $\tau_{\rm fast,exp}$ compared with the estimated values based on kinetic theory $\tau_{\rm fast,kinetic}$, and a simplified Monte-Carlo study $\tau_{\rm fast,MC}$. The MC studies do not include sidewall losses so no value can be calculated for measurements without remover.}
\label{tab:tauFast}
\begin{tabular}{c c c c c c c }
\hline\hline
remover &     $\tau_{\rm fast,exp}$ & $\tau_{\rm fast,kinetic}$ & $\tau_{\rm fast,MC}$ \\ \hline
no       &        $(80 \pm 16)\un{s}$ &      $56\un{s}$          &  --   \\
80cm PE  &   $(30 \pm 7)\un{s}$ &       $48\un{s}$          & 27\un{s}    \\
65cm PE &    $(37 \pm 7)\un{s}$ &       $34\un{s}$          & 28\un{s}    \\\hline\hline
\end{tabular}
\end{center}
\end{table}

A summary of the calculated $\tau_{\rm fast}$ based on kinetic theory and the simplified MC model are shown in Table~\ref{tab:tauFast}. These values compare well with the experimentally observed $\tau_{\rm fast}$. This demonstrates that we have good understanding of the cleaning process.

The effectiveness of cleaning is not determined by how short $\tau_{\rm fast}$ is; rather, it is determined by how many UCNs with $E_{\rm tot}>U_{\rm trap}$ remain after cleaning. This is primarily decided by how short $\tau_{\rm clean}(U_{\rm trap})$ is, and at energies just above $U_{\rm trap}$. To study this, we calculate the fraction $N(t_{\rm hold},U_{\rm trap}<E_{\rm tot}<U_{\rm trap}+\Delta E)/N_0(E_{\rm tot}<U_{\rm rm})$. $U_{\rm rm}$ is used in the denominator because having a lower $U_{\rm rm}$, while providing a shorter $\tau_{\rm clean}(U_{\rm trap})$, causes a greater loss of well-trapped UCNs. The measured UCN spectrum from Eq.~\ref{eq:integralSpectrumFit} is assumed.

\begin{figure}[tb!]
\begin{center}
\includegraphics[width=3.2in]{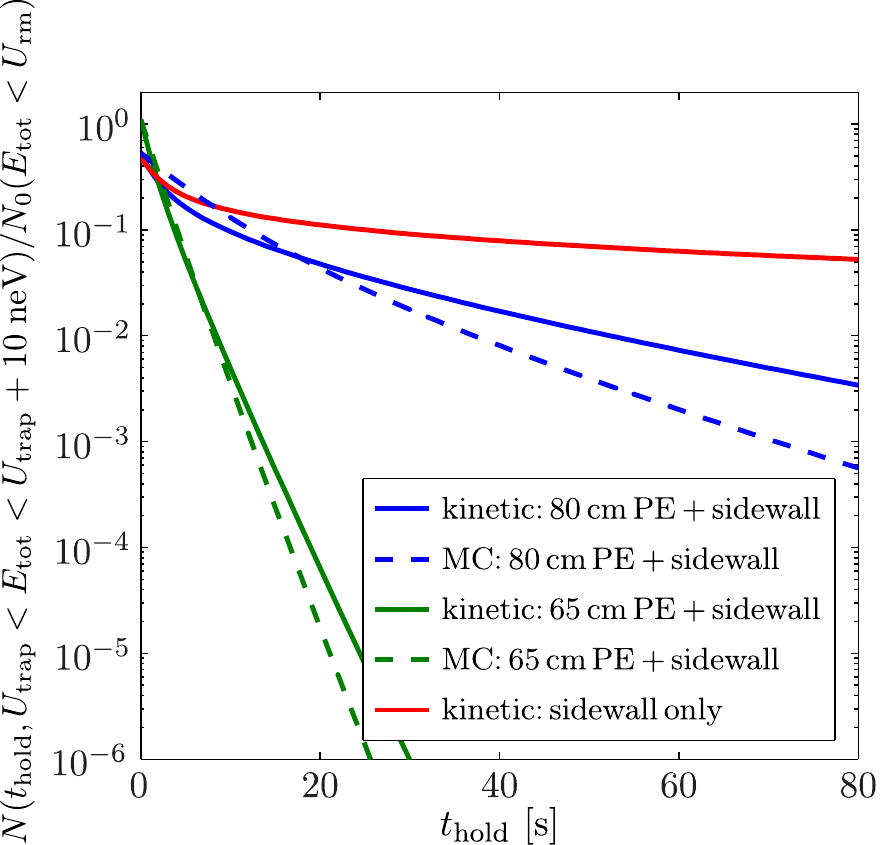}
\caption{(Color online) The calculated number of the UCNs with $U_{\rm trap}<E_{\rm tot}<U_{\rm trap}+10\un{neV}$ remaining during the cleaning process (i.e. $t_{\rm hold} \leq 80\un{s}$) normalized to the initial number of UCNs with $E_{\rm tot} < U_{\rm rm}$. The latter is approximately constant to $\sim 10\%$ during cleaning because of the long $\tau_{\rm tot}$ for well-trapped UCNs ($E_{\rm tot}<U_{\rm trap}$). This is done for all three remover configurations using the measured $E_{\rm tot}^{2.5}$ UCN spectrum and based on the kinetic theory and a simplified MC study. This plot is used the estimate the fraction of above-threshold UCNs remaining after cleaning to study the effectiveness of a cleaning procedure.}
\label{fig:effectiveCleaning}
\end{center}
\end{figure}

An analysis using $\Delta E = 10\un{neV}$ is shown in Fig.~\ref{fig:effectiveCleaning}. It can be seen that when no remover is used, $\sim 5\%$ of the UCNs remaining after cleaning ($t_{\rm hold} = 80\un{s}$) are above-threshold UCNs  ($E_{\rm tot}> U_{\rm trap}$). This fraction decreases by an order of magnitude for the 80\un{cm} remover measurement and decreases further by another factor of $10^{12}$ for the 65\un{cm} remover measurement. By scanning the value of $\Delta E$, it was found that all the above-threshold UCNs remaining after 80\un{s} have $E_{\rm tot} < U_{\rm trap} + 5\un{neV}$ for all three configurations. 

%Implication for full magnetic trap

\section{Extraction of the neutron lifetime \label{sec:lifetime}}

Besides the decay of the free neutron, there are other mechanisms in the trap that cause UCNs loss. We denote $\tau_{\rm tot}$ as the total storage time constant, which is given by:
\begin{equation}
\tau_{\rm tot}^{-1}(E_{\rm tot}) \equiv \tau_{\rm n}^{-1} + \tau_{\rm walls}^{-1}(E_{\rm tot}) + \tau_{\rm gas}^{-1} + \tau_{\rm depol}^{-1}(E_{\rm tot}) \;, 
\end{equation}
where $\tau_{\rm depol}^{-1}$ comes from loss due to depolarization, which, as will be shown in Sec.~\ref{sec:depolarization}, is suppressed to below the statistical sensitivity of these measurements; $\tau_{\rm gas}^{-1}$ is from interactions with residual gas in the vacuum; and $\tau_{\rm walls}^{-1}(E_{\rm tot})$ comes from interactions of UCNs with the Fomblin grease at the bottom and sidewall of the trap. This assumes that the cleaning process has been sufficient efficient. Therefore, if $\tau_{\rm walls}(E_{\rm tot})$ and $\tau_{\rm gas}$ can be corrected for, $\tau_{\rm n}$ be extracted.

Due to the energy-dependence of $\tau_{\rm tot}(E_{\rm tot})$, the total number of well-trapped UCNs in the trap is given by:
\begin{eqnarray} \label{eq:decaycurveintegral}
N(t_{\rm hold},E_{\rm tot}<U_{\rm trap}) \nonumber \\*= \int^{U_{\rm rm}}_0 \!n_0(E_{\rm tot})\,\exp\left[\frac{-t}{\tau_{\rm tot}(E_{\rm tot})}\right] {\rm d}E_{\rm tot} \;.
\end{eqnarray}
By assuming kinetic theory, $\tau_{\rm walls}(E_{\rm tot})$ can be calculated using Eq.~\ref{eq:tauCleanUCNgas}. $V_{\rm eff}(E_{\rm tot})$ for $E_{\rm tot}<U_{\rm trap}$ when the remover is removed is shown in Fig.~\ref{fig:effectiveVolume}. The $A_{\rm eff\,loss}(E_{\rm tot})/f_{\rm Fomblin}$ from Eq.~\ref{eq:Aeffloss} for the Fomblin grease at the bottom and sidewall, as well as $\tau_{\rm walls}^{-1}(E_{\rm tot})/f_{\rm Fomblin}$, are shown in Fig.~\ref{fig:tauWalls}. 

Here we provide estimates of the systematic uncertainties involved in deducing $\tau_{\rm walls}(E_{\rm tot})$ and $\tau_{\rm gas}$:
\begin{itemize}
  \item \emph{$f_{\rm Fomblin}$}: A measurement using the same Fomblin grease in our experiment yielded $f_{\rm Fomblin} = 1.85 \pm 0.10 \times 10^{-5}$ \cite{Richardson1991}. At room temperature, there are around 5--6 measurements made with Fomblin oil that lie in the range between $f_{\rm Fomblin} = 6.7 \times 10^{-6} \text{ to } 2.5 \times 10^{-5}$ (see \cite{Pokotilovski2008} for a review). $f_{\rm Fomblin}$ is temperature-dependent, changing by $\sim 3\un{\%\,K^{-1}}$ at room temperature. To be conservative, we will use a value of $f_{\rm Fomblin} =(1.6 \pm 0.9)\times 10^{-5}$. This covers the range of values measured for both Fomblin grease and oil and takes into account temperature variations of $\sim \pm 10{\rm ^\circ C}$.
  \item \emph{$U_{\rm Fomblin}$}: a measurement with Fomblin grease produced a value of $107.5^{+1}_{-2}\un{neV}$ \cite{Golub1991}. Fomblin oil, which has been much more studied, has $U_{\rm Fomblin} = 106.5\un{neV}$ \cite{Lamoreaux2002}. The two agree within the experimental limits. A conservative value of $U_{\rm Fomblin} = 107 \pm 5 \un{neV}$ will be used.
    \item \emph{trap volume}: The bottom of the trap is completely flat; the piston is recessed from the valve seat by 7\un{mm}, forming a volume of $20\un{cm^3}$. There are also variations in the physical diameter of the magnet bore as well. A conservative systematic error in $V_{\rm eff}(E_{\rm tot})$ of $\pm 100\un{cm^3}$ for all $E_{\rm tot}$ will be used.
  \item \emph{UCN spectrum}: The differential spectrum $n_0(E_{\rm tot}) \propto E_{\rm tot}^{(2.50 \pm 0.16)}$ was extracted from measurements of the integral spectrum using four different $h_{\rm PE}$. A conservative error of $n_0(E_{\rm tot}) \propto E_{\rm tot}^{2.5 \pm 0.5}$ is used our in analysis.
  \item \emph{magnet strength}: The variation in the magnets' magnetization (or magnetic field strength at the walls) was measured to be $< 5\%$ \cite{Leung2013a}. A conservative estimate of a $\pm 10\%$ variation will be used.
  \item \emph{$\tau_{\rm gas}^{-1}$}: the loss rate of trapped UCNs with residual gas is given by: $\tau_{\rm gas}^{-1} = \Sigma_i\, n_{{\rm gas},i}\,\sigma_{i} \, v_{{\rm gas},i}$, where: $n_{{\rm gas},i}$ is the number density of the $i$th-species of gas in the vacuum; $v_{{\rm gas},i}$ is the speed of the $i$th-species gas molecule (assuming $v_{\rm gas} \gg v$); and $\sigma_{i}$ is the loss cross-section, which can come from absorption and scattering. For 100\% hydrogen in our $5\times10^{-5}\un{mbar}$ vacuum, we expect $\tau^{-1}_{\rm gas} = 6\times10^{-6}\un{s^{-1}}$ \cite{Leung2013a}. If the residual gas is 100\% H2O vapor then we expected $\tau^{-1}_{\rm gas} = 4\times10^{-6}\un{s^{-1}}$. A value for $P\,\tau_{\rm gas} = (9.5 \pm 0.5)\un{mbar\, s}$ at 113\un{K} was measured in \cite{Serebrov2008} by deliberately worsening the vacuum from $5\times10^{-6}\un{mbar}$ to $8 \times 10^{-4}\un{mbar}$. Using this value, correcting for the temperature difference with our experiment, we expect $\tau^{-1}_{\rm gas} = 3\times10^{-6}\un{s^{-1}}$. For our analysis, we will use a conservative value of $\tau^{-1}_{\rm gas} = (4.5 \pm 1.5 )\times10^{-6}\un{s^{-1}}$, which covers both the above values.
\end{itemize}

Due to limited statistics, only one single exponential decay constant can be fitted from a decay curve, which we will denote as $\tau_{\rm fit}$. To show the relation between $\tau_{\rm fit}$, which comes from a Levenberg-Marquardt non-linear regression data-fitting algorithm, and $\tau_{\rm tot}(E_{\rm tot})$ we perform data fitting on simulated data. First, let us define two quantities, $\bar{\tau}_{\rm walls}^{-1}$ and $\bar{\tau}_{\rm tot}^{-1}$, given by:
\begin{equation} \label{eq:bartauwalls}
\bar{\tau}_{\rm walls}^{-1} \equiv \frac{ \int^{U_{\rm rm}}_0 n_0(E_{\rm tot})\,\tau_{\rm walls}^{-1}(E_{\rm tot}) \,{\rm d}E_{\rm tot}}{ \int^{U_{\rm rm}}_0 n_0(E_{\rm tot})\,{\rm d}E_{\rm tot}}
\end{equation}
and
\begin{equation} \label{eq:bartautot}
\bar{\tau}_{\rm tot}^{-1} \equiv \tau_{\rm n}^{-1} + \tau_{\rm gas}^{-1} + \bar{\tau}_{\rm walls}^{-1}\;.
\end{equation}
These quantities are not explicitly $E_{\rm tot}$ dependent. We will assert that:
\begin{equation}
\label{eq:fittedTauCalculatedTau}
\tau_{\rm fit} \approx \bar{\tau}_{\rm tot}\;.
\end{equation}
We will show this approximation holds within the statistical errors of the extracted $\tau_{\rm fit}$ values.

$\tau_{\rm walls}(E_{\rm tot})$ and $n_0(E_{\rm tot})$ were varied according to the conservatively estimated systematic errors described previously. $N(t_{\rm hold}, E_{\rm tot}< U_{\rm trap})$ data points were generated from Eq.~\ref{eq:decaycurveintegral} with statistical errors in each point similar to those in our measurements. These points were fitted with a single exponential decay $\tau_{\rm fit}$. The spacing between the simulated points and the number of them were also varied. The extracted $\tau_{\rm fit}$ were compared with the calculated $\bar\tau_{\rm tot}$ from Eq.~\ref{eq:bartautot}. It was found that the two values always agreed to within the statistical error of the $\tau_{\rm fit}$ values.

To deliberately see the approximation breaking down, we set the statistical error of the generated $N(t_{\rm hold}, E_{\rm tot}< U_{\rm trap})$ data points to zero. It was found that the difference between $\tau_{\rm fit}$, which now has negligible error, and $\bar\tau_{\rm tot}$ to be $\sim 0.1\un{s}$. We then increased the size of $\tau^{-1}_{\rm walls}(E_{\rm tot})$ by a factor of 10, and now the difference between $\tau_{\rm fit}$ and $\bar\tau_{\rm tot}$ becomes $\sim 1\un{s}$. Increasing $\tau^{-1}_{\rm walls}(E_{\rm tot})$ further, we saw that we require two-orders-of-magnitude larger wall loss before the break down of the approximation has an effect comparable to the statistical precision of our measurements.

\begin{figure}[tb!]
\begin{center}
\includegraphics[width=3in]{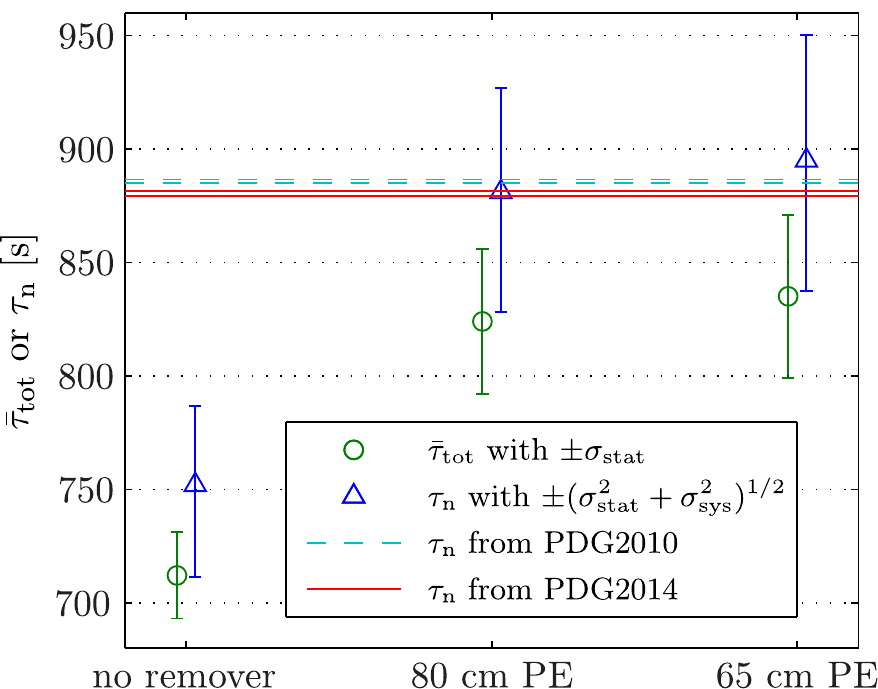}
\end{center}
\caption{(Color online) The observed total storage time constants for the three remover configurations for $t_{\rm hold}> 80\un{s}$ (i.e. after cleaning) with statistical error bars ($\sigma_{\rm stat}$). The extracted neutron lifetimes $\tau_{\rm n}$, assuming effective cleaning, are plotted next to these values. The error bars have the statistical error and the total systematic error combined in quadrature [$(\sigma_{\rm stat}^2 + \sigma_{\rm sys}^2)^{1/2}$]. These values are compared with the Particle Data Group (PDG) values \cite{Nakamura2010, Olive2014}.}
\label{fig:tauExtracted}
\end{figure}

Now that we have verified Eq.~\ref{eq:fittedTauCalculatedTau}, we can extract $\tau_{\rm n}$ from Eqs.~\ref{eq:bartauwalls} and \ref{eq:bartautot}. We will set $\tau_{\rm fit}$ to be the first $\tau_{\rm 1\text{-}exp}$ value when the starting $t_{\rm hold} > 80\un{s} = t_{\rm clean}$ (see Fig.~\ref{fig:singleExpFits}) for all three remover configurations \footnote{In \cite{Leung2013a} the value for $\tau_{\rm fit}$ used were $\tau_{\rm slow}$ and thus some of the numbers in the subsequent analysis are slightly different; however, the conclusions remain the same.}. Table~\ref{tab:tauExtraction} shows the corrections required for extracting $\tau_{\rm n}$ as well as the uncertainty contribution from each parameter, as well as the total systematic uncertainty from a linear sum. These values are plotted in Fig.~\ref{fig:tauExtracted} and compared with Particle Data Group (PDG) values \cite{Nakamura2010, Olive2014}.

\begin{table}[htdp]
\small
\caption{The extracted $\tau_{\rm n}$ values from the measurements with the three remover configurations after correcting $\bar{\tau}^{-1}_{\rm tot}$ for $\bar{\tau}^{-1}_{\rm walls}$ and $\tau^{-1}_{\rm gas}$. The systematic error contribution of each effect explained in the text are also calculated. The bold font is used to emphasize values that are in units of time.}
\begin{center}
\begin{tabular}{c c  c c c c}
\hline\hline
					& no remover & 80 cm PE & 65 cm PE \\ \hline
$\bar{\tau}_{\rm tot}$		&{\bf 712} & {\bf 824} & {\bf  835} & [s]\\
$\bar{\tau}^{-1}_{\rm tot}$	& 1.405 & 1.213  & 1.197 & $[\times 10^{-3}\un{s^{-1}}]$ \\ 
$\bar{\tau}^{-1}_{\rm walls}$ & 7.0 & 7.4 & 7.5 & $[\times 10^{-5}\un{s^{-1}}]$ \\ 
${\tau}^{-1}_{\rm gas}$  & 0.45 & 0.45  & 0.45 & $[\times 10^{-5}\un{s^{-1}}]$ \\ 
$\tau^{-1}_{\rm n}$     & 1.331 & 1.135 & 1.118 & $[\times 10^{-3}\un{s^{-1}}]$ \\
$\tau_{\rm n}$ & {\bf 752} & {\bf 881} & {\bf 895} & [s] \\ \hline
\multicolumn{2}{l}{ \emph{statistical uncertainty}:} \\
$\sigma_{\rm stat} \text{ in } \tau_{\rm n}$ & {\bf $\pm$19} & {\bf $\pm$32} & {\bf $\pm$36} & [s] \\  \hline
\multicolumn{2}{l}{ \emph{systematic uncertainties}:} \\
$f_{\rm Fomblin}$   & +3.9/-4.1 & +3.9/-4.3 & +4.3/-4.3 & $[\times 10^{-5}\un{s^{-1}}]$\\
$U_{\rm Fomblin}$ & +1.9/-2.4 & +2.5/-2.7 & +3.1/-2.3 & $[\times 10^{-6}\un{s^{-1}}]$ \\
$\tau_{\rm gas}^{-1}$  & $\pm$1.5 & $\pm$1.5 & $\pm$1.5 & $[\times 10^{-6}\un{s^{-1}}]$\\
{\scriptsize trap volume} & +0.02/-7.8 & +0.4/-6.1 & +1.3/-5.4 &  $[\times 10^{-6}\un{s^{-1}}]$  \\
{\scriptsize magnet strength}  & +0.1/-4.7 & +2.0/-6.1 & +3.6/-6.2 & $[\times 10^{-7}\un{s^{-1}}]$ \\
{\scriptsize UCN spectrum} & +2.4/-11 & +4.4/-11 & +9.9/-7.1 &  $[\times 10^{-7}\un{s^{-1}}]$ \\ \hline 
\multicolumn{2}{l}{ \emph{total systematic uncertainty}:} \\
$\sigma_{\rm sys}\text{ in } \tau^{-1}_{\rm n}$ & +4.3/-5.4 & +4.4/-5.5 & +5.0/-5.4 & $[\times 10^{-5}\un{s^{-1}}]$  \\
$\sigma_{\rm sys}\text{ in } \tau_{\rm n}$& {\bf +29/-36} & {\bf +33/-42} & {\bf +42/-45} & [s] \\ \hline\hline
\end{tabular}
\end{center}
\label{tab:tauExtraction}
\end{table}

\section{Depolarization loss studies \label{sec:depolarization}}

An experiment performed in 2009 scanning the bias field solenoid current is used to demonstrate that the loss due to depolarization is less than the sensitivity of the extracted $\tau_{\rm n}$. In these earlier measurements, the permanent magnet array were used with the long axis of the trap at a 30$^\circ$ angle relative to the horizontal. The same copper valve seat, PTFE piston valve, and bias field solenoid were used. The angled trap was an earlier tested scheme of using the gravitational axis to break the symmetry of the trapping potential to improve cleaning.

\begin{figure}[tb!]
\begin{center}
\includegraphics[width=3in]{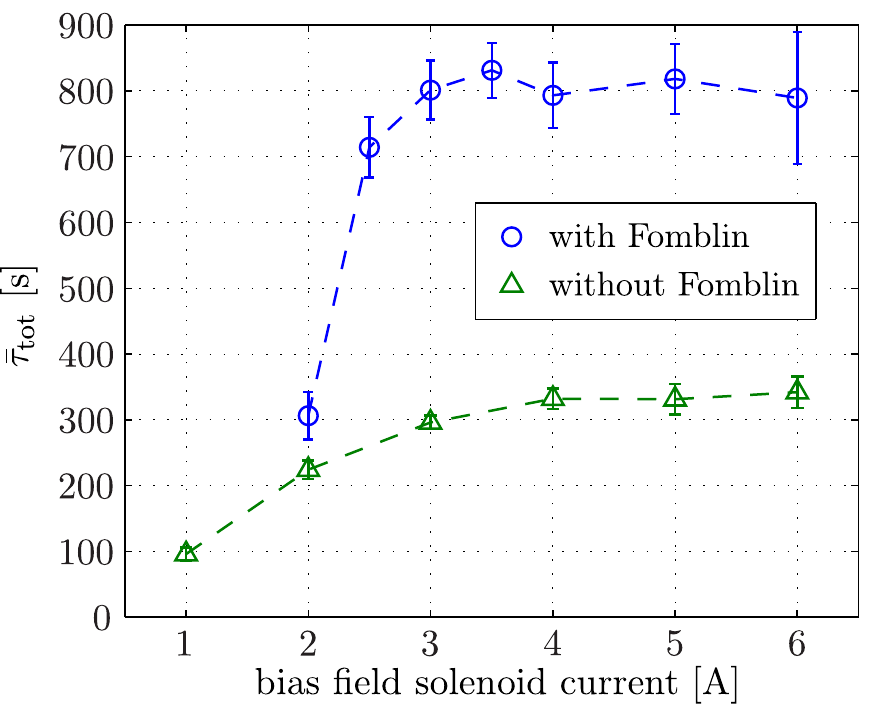}
\end{center}
\caption{The total storage time constant $\bar{\tau}_{\rm tot}$ when scanning the bias field solenoid current. The measurements were made in earlier angled-trap experiments (see text) before Fomblin grease was applied on the valve body and valve seat (``without Fomblin'') and after Fomblin grease was applied (``with Fomblin''). The dashed lines are used to guide the eye only.}
\label{fig:depolarizationTau}
\end{figure}

Two sets of measurements were made: one before Fomblin grease was used to coat the valve body and valve seat, and the other after (hence the shorter $\bar{\tau}_{\rm tot}$ of the former). In both sets of data, Fomblin grease was on the sidewall of the magnets 8\un{cm} in from the lower end. No UCN remover inside the trap was used in these experiments. Instead, an external UCN gravitational spectrometer/cleaner (described in \cite{Richardson1991}) was used outside the trap for pre-cleaning of the UCN spectrum before loading into the trap. The results are plotted in Fig.~\ref{fig:depolarizationTau}. The extracted $\tau_{\rm depol}$ values are shown in Table~\ref{tab:tauDepol}, calculated assuming $\tau_{\rm depol}$ at 5\un{A} is sufficiently large as to not cause a statistically significant shift in $\bar{\tau}_{\rm tot}$.

\begin{table}
\begin{center}
\caption{The depolarization loss time constant $\tau_{\rm depol}$ extracted from the $\bar{\tau}_{\rm tot}$ measurements when scanning the bias field solenoid current. Assumes that $\tau_{\rm depol}$ at 5\un{A} is large. The values for currents $>3.5\un{A}$ are not shown due to their large statistical errors.}
\label{tab:tauDepol}
\begin{tabular}{c c c }
\hline\hline
 {\small solenoid}  & ${\tau}_{\rm depol}$ [s] & ${\tau}_{\rm depol}$ [s] \\ 
{\small current} [A]    &     {\small with Fomblin} & {\small without Fomblin} \\ \hline
%4A  &    $(30,000 \pm 70,000)\un{s}$     &  $(-100,000 \pm 3,100,000)\un{s}$   \\\hline
%3.5A  & $(-50,000 \pm 270,000)\un{s}$  &   --  \\\hline
3.0  &    $40\;000 \pm 160\;000$   &  $2\;800 \pm 1\;800$ \\
2.5  &  $5\;600 \pm 3\;800$           &  --    \\
2.0  &  $490 \pm 90$                   &   $690 \pm 170$    \\
1.0  &  --                                      &   $135 \pm 20$    \\\hline\hline
\end{tabular}
\end{center}
\end{table}

The region of lowest $|\vec{B}|$ occurs at the central axis of the trap. This is the region that is most problematic for the depolarization of low-field seeking UCNs. A 3-axis hall probe with a Lakeshore 460 3-channel gauss-meter was inserted up to 40\un{cm} into the trap from both ends to measure the field along the central axis \cite{Fraval2009}. With no current in the coil, features with amplitudes between 1$\,$--$\,$3\un{mT} with components both parallel and anti-parallel to the long axis, and with a spacing approximately 10\un{cm} apart, were measured along the central axis. This was suspected to be caused by differences in the magnetization of the permanent magnets between modules. The axial field provided by the solenoid at 3\un{A} was measured to be 4\un{mT}, which matches calculations. Therefore, we conclude that at 3\un{A}, the minimum field inside the trapping region is $\sim 1\un{mT}$. At 4\un{A}, the current used in the measurements described in the previous two sections, the minimum field is $\sim2\un{mT}$.

\section{Conclusion \label{sec:conclusion}}

The results from the 1st phase measurements using the HOPE UCN trap were presented in this paper. In this configuration the octupole magnet array is aligned vertically, a Fomblin grease coated piston valve is used at the bottom of the trap, and a flat movable polyethylene UCN remover inserted from the top. In Sec.~\ref{sec:experimentSetup} the effective volume of the trap $V_{\rm eff}(E_{\rm tot})$ is calculated. Also in this section, a quantity defined called the effective loss areas $A_{\rm eff\,loss}(E_{\rm tot})$ is defined in this paper. This was calculated for the Fomblin grease at the bottom and sidewall, as well as for the PE remover and epoxy on the sidewall.

Using the PE remover at four heights the differential UCN spectrum $N_0(E_{\rm tot})$ loaded into the trap was measured (Fig.~\ref{fig:effectiveVolume}). Three detailed UCN storage curves were made for without the remover and with the remover at 80\un{cm} and 65\un{cm} from the bottom of the trap. When using the remover, it was lowered in position for $80\un{s}$ of cleaning time before being raised up again. The piston valve at the bottom of the trap allowed us to empty the UCNs with a time constant of $\sim 2\un{s}$. This provided a unique opportunity to study the time behavior of the UCNs during short holding times $t_{\rm hold}$, including during cleaning. The decay curves were fitted with a sum of two exponential decays, with time constants $\tau_{\rm fast}$ and $\tau_{\rm slow}$ (Fig.~\ref{fig:storageCurve80cm} and Table~\ref{tab:twoExpDecays}), as well as with a single exponential while scanning the starting $t_{\rm hold}$ of the fit (Fig.~\ref{fig:singleExpFits}).

To understand the cleaning process, calculations assuming UCN kinetic theory using $V_{\rm eff}(E_{\rm tot})$, $A_{\rm eff\,loss}(E_{\rm tot})$ and $n_0(E_{\rm tot})$ were made (see Sec.~\ref{sec:cleaning}). A MC calculation of a simplified geometry was also done for a comparison. $\tau_{\rm fast}$, the parameter describing the UCNs during cleaning, was found to agree well with these calculations (Table~\ref{tab:tauFast}). This demonstrates that we have good understanding of the cleaning process.

The kinetic theory calculations and the MC simulations were also used to predict the fraction of above-threshold UCNs ($E_{\rm tot}>U_{\rm trap}$) remaining after cleaning compared to the number of remaining well-trapped UCNs ($E_{\rm tot}<U_{\rm rm}$). This was found to be $\sim 5\%$ for no remover, $\sim 0.5\%$ for 80\un{cm} remover and $\sim 10^{-12} \%$ for the 65\un{cm} remover measurements, with all the surviving above-threshold UCNs having $E_{\rm tot}$ less than $U_{\rm trap}+5\un{neV}$.

The fitted total storage time $\tau_{\rm fit}$ extracted from a measurement is given by fitting the number of remaining UCNs $N(t_{\rm hold})$ with a single exponential decay after the cleaning process. For the 65\un{cm} remover measurement this value is $(835\pm36)\un{s}$ and for the 80\un{cm} remover it is $(824\pm32)\un{s}$. When no remover is used the value extracted is $(712\pm19)\un{s}$ by fitting only for $t_{\rm hold} > 80\un{s}$, to match that done for the other two measurements. In Sec.~\ref{sec:lifetime}, it was shown by fitting simulated data points that $\tau_{\rm fit} \approx \bar{\tau}_{\rm tot} = (\tau_{\rm n}^{-1} + \tau_{\rm gas}^{-1} + \bar{\tau}_{\rm walls}^{-1})^{-1}$.  $\bar{\tau}_{\rm walls}^{-1}$ is the $E_{\rm tot}$ averaged loss rate that can be calculated with the measured $n_0(E_{\rm tot})$, $V_{\rm eff}(E_{\rm tot})$ and $A_{\rm eff\,loss}(E_{\rm tot})$ of the Fomblin grease at the bottom and sidewall of the trap. Therefore, we have shown that $\tau_{\rm n}$ can be extracted from $\tau_{\rm fit}$. 

The extracted $\tau_{\rm n}$ and the statistical and the conservative systematic uncertainties (described in Sec.~\ref{sec:lifetime}) are summarized in Table~\ref{tab:tauExtraction}. A value of $(895 \pm 36_{\rm \, stat}\, \vspace{0mm}^{+42}_{-45} \vspace{0mm}_{\rm \,sys})\un{s}$ was extracted for the 65\un{cm} remover measurement, $(881 \pm 32_{\rm \,stat}\, \vspace{0mm}^{+33}_{-42}\vspace{0mm}_{\rm \,sys})\un{s}$ for the 80\un{cm} remover, and $(712 \pm 19_{\rm \, stat}\,\vspace{0mm}^{+29}_{-36} \vspace{0mm}_{\rm \, sys})\un{s}$ for without the remover. The former two values agree with the PDG values (see Fig.~\ref{fig:tauExtracted}), while the latter from without the remover exhibits a $\Delta \tau_{\rm n} = (128\,\vspace{0mm}^{+35}_{-41})\un{s}$ disagreement (i.e. approximately 3.5 $\sigma$) when combining the statistical and systematic uncertainties in quadrature, clearly showing the effects of poor cleaning. The two values from the measurements using the remover can be combined in a weighted average, with the larger of the asymmetric systematic errors combined with the statistical error in quadrature, to produce a value of $\tau_{\rm n} = (887 \pm 39) \un{s}$ determined from these 1st phase measurements.

By using generated data with the same holding time and statistical errors as the no remover measurement, the possible $\Delta \tau_{\rm n}$ caused by above-threshold UCNs can be estimated. It was found that if the above-threshold UCNs have a total storage time constant anywhere between $\sim150\text{--}450\un{s}$, a 5\% fraction in the no remover measurement can produce $\Delta \tau_{\rm n} = (21 \pm 28)\un{s}$, smaller than the shift observed. It is more likely that the fraction of above-threshold UCNs present was $\sim 10\text{--}15\%$. This is still fairly good agreement for a population of UCNs that are difficult to understand, especially since the Fomblin grease at the bottom of the trap was not at an uniform height, as it was applied by hand, and only kinetic theory calculations were made for the no absorber configuration.

The tiny fraction of above-threshold UCNs remaining after cleaning for the 65\un{cm} remover measurement suggests that the remover height and also the cleaning time can be further optimized in the experiment. For instance, if a 75\un{cm} remover is used instead while keeping the 80\un{s} cleaning time, $\Delta \tau_{\rm n} \sim 10^{-3}\un{s}$ is expected, better than two orders of magnitude below requirement. The percentage of well-trapped UCNs remaining for the 75\un{cm} remover compared to when no absorber is used would be 70\%. For the 85\un{cm} and 65\un{cm} remover, this percentage is 90\% and 45\%. Furthermore, these numbers are dependent on the UCN spectrum: for a softer spectrum the remaining well-trapped percentage would increase.

In Sec.~\ref{sec:depolarization} the effects of depolarization loss of UCNs were studied in measurements scanning the bias field solenoid current. This showed that the minimum 2\un{mT} bias field used for the measurements does not cause a statistically significant loss within the uncertainties of the measurements.

\section{Future \label{sec:future}}

\begin{figure}[tb!]
\begin{center}
\includegraphics[width=3.0in]{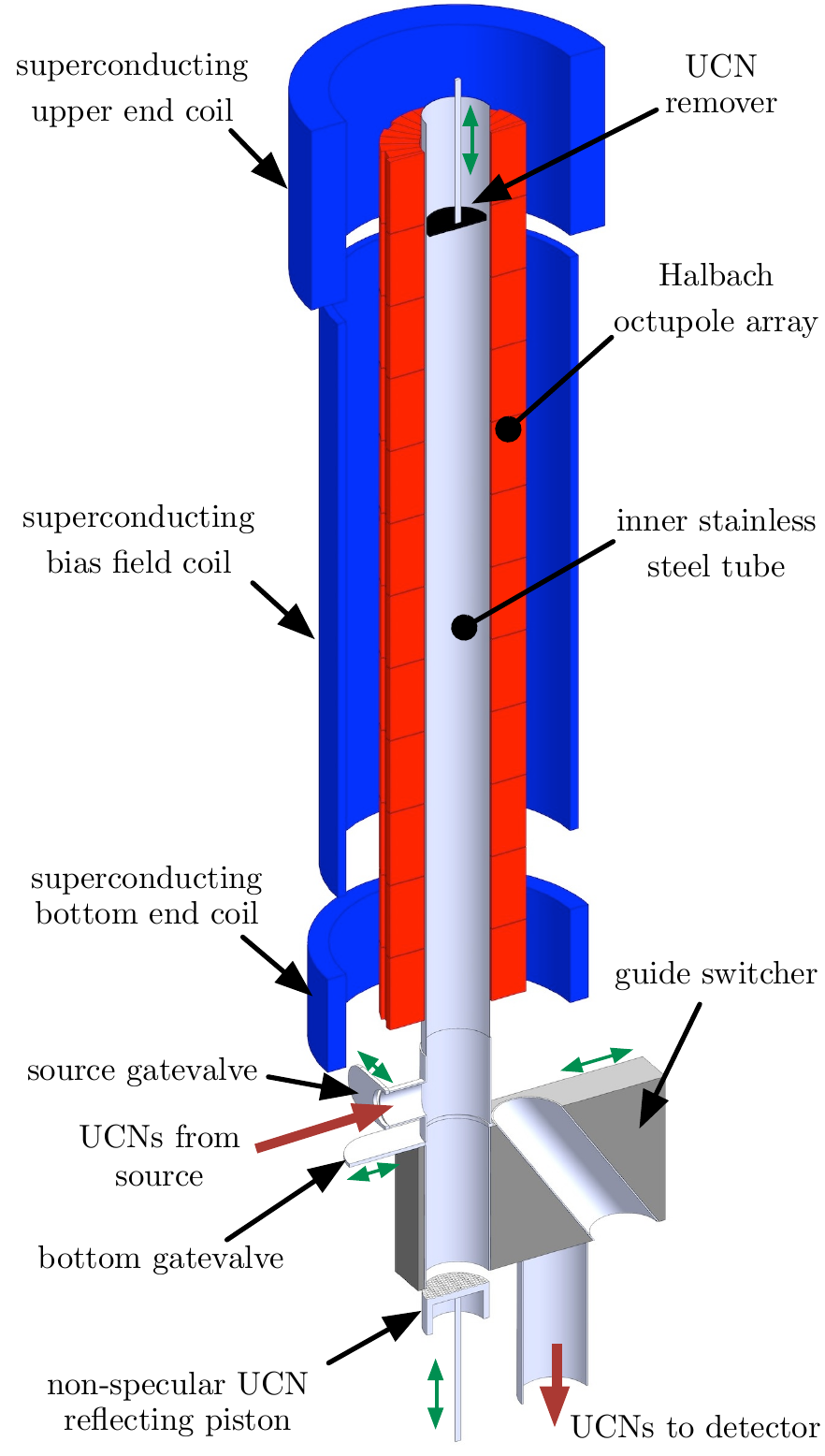}
\end{center}
\caption{(Color online) Schematic of the next phase measurements of the HOPE project with full 3D magnetic trapping. The movable components are shown with the green arrows and nominal ``flow'' of UCNs into and out of the shown parts of the setup are indicated by the red arrows. The superconducting upper end coil is only used if focussing of the charged neutron decay products onto a detector is required (not discussed in this paper).}
\label{fig:3DtrapSetup}
\end{figure}

These 1st phase measurements played a crucial role in demonstrating a successful spectral cleaning scheme, which guided the trap design to be vertical for the next phase measurements of the HOPE project. These measurements will involve a full 3D magnetic trap by adding a bottom superconducting end coil and a superconducting bias field solenoid. There will also be a thin-walled stainless steel tube inserted inside the magnet inner bore (Fig.~\ref{fig:3DtrapSetup}).

We plan on employing the same cleaning technique demonstrated here by using a movable non-specular reflecting piston that can be inserted into the trap from the bottom. The end coil is ramped up after the piston is inserted, so that the UCNs are not heated by the increasing magnetic field. Having the stainless steel on the sidewall does not reduce the cleaning time in the trap (see Fig.~\ref{fig:cleaningAreasTimes}). This cleaning scheme also does not cause a large loss of well-trapped UCNs because even with the Fomblin grease at the bottom of the trap $\bar{\tau}_{\rm tot} \approx 820\un{s}$. This technique also has the advantage that when the reflector and remover is retracted after cleaning, UCNs are not heated from doppler reflections. A more detailed discussion of other cleaning schemes considered is given in \cite{Leung2013a}.

Besides the cleaning scheme, which has been discussed in detail already, the bias field used for suppressing UCN depolarization can be as large as 0.3\un{T} because of the superconducting solenoid. To indirectly study the effect of depolarization, the bias field strength can be scanned to see its effect on $\bar{\tau}_{\rm tot}$.

Indirect depolarization studies done this way are only sensitive to depolarization loss in the low field regions of the trap, and not by field imperfections near the wall. In the 3D magnetic field setup, the bottom of the trap will be coupled to a UCN detector so that depolarized UCNs can be counted during the holding time. This is allowed by the inner stainless steel tube, which has $U_{\rm SS} = 185\un{neV}$. Live depolarization monitoring will let us set an experimentally determined upper limit of the depolarization loss rate. Another advantage offered by the tube is that it separates the high outgassing hydrocarbon-based epoxy used on the magnet walls from the vacuum space of the UCNs. This will reduce loss caused by residual gas. In 2014 commissioning measurements with the superconducting coils were performed \cite{Rosenau2015}. 

The HOPE magnetic trap's volume is relatively small, a $2\un{L}$ physical volume only, however, the experimental design as discussed above offers a great control of systematics. The $\sim 30\un{s}$ statistical precision of each $\tau_{\rm n}$ measurement in this paper took approximately 2 days on the PF2-TES beam, which is $\sim 20$ times weaker than the other UCN beams on the PF2 source. This suggests \footnote{Includes optimization of the holding times used as well as a decrease in $U_{\rm trap}$ when the superconducting coils and stainless steel tube is used.} that a statistical precision of $\sim 5\un{s}$ can be reached at other ports of PF2 for a 10-live-day measurement, allowing time for systematic studies within a 50 day cycle. 

A better scheme is that planned use of the HOPE magnetic trap with the SUN2 high-density superfluid helium UCN source at the ILL \cite{Leung2016, Piegsa2014, Zimmer2011}. This source is optimized for filling small volumes and offers a softer UCN spectrum, which is better suited for the HOPE magnetic trap. With this configuration a robust $\sim$1-s-precision measurement with excellent control of systematics, much needed for the neutron $\beta$-decay community, can be made. Furthermore, for sub-1-s-precision measurements, the future SuperSUN source \cite{Zimmer2015} can be used.

\section{Acknowledgements \label{sec:acknowledgements}}
We would like to thank ILL's PF2 and Nuclear and Particle Physics group technicians, T.~Brenner, D.~Berruyer, and R.~Bebb, for their support. We would also like to acknowledge the help of our summer \emph{stagiaires}, K.~Fraval and O.~Roberts, as well as F.~Piegsa for their help during the experiments. Support from the Deutsche Forschungsgemeinschaft (DFG) is gratefully acknowledged. KL's time for data analysis and preparation of this paper was supported in part by the US Department of Energy under Grant No. DE-FG02-97ER41042.

\bibliographystyle{apsrev4-1}
\footnotesize
\bibliography{HOPEpaper}

\end{document}